\documentclass[reqno,12pt]{amsart}
\usepackage{amsfonts}
\usepackage{amssymb}
\usepackage{graphicx}
\usepackage{fullpage}
\usepackage{url}

\begin{document}

\newtheorem{theorem}{Theorem}
\newtheorem{lemma}{Lemma}
\newtheorem{proposition}{Proposition}
\numberwithin{equation}{section}

%my macros for LaTeX fixes

% refs:

%put `Eq.( )' around an equation ref in latex
\def\Eqref#1{Eq.~(\ref{#1})}
\def\Eqrefs#1#2{Eqs.~(\ref{#1}) and~(\ref{#2})}
\def\Eqsref#1#2{Eqs.~(\ref{#1}) to~(\ref{#2})}
\def\Sysref#1#2{Eqs. (\ref{#1})--(\ref{#2})}

%put `Fig.' before a figure ref in latex
\def\figref#1{Fig.~\ref{#1}}

%put `Sec. ' before a section ref in latex
\def\secref#1{Sec.~\ref{#1}}
\def\secrefs#1#2{Sec.~\ref{#1} and~\ref{#2}}

%put `App. ' before an appendix ref in latex
\def\appref#1{Appendix~\ref{#1}}

%put `Ref. ' before a bibitem ref in latex
\def\Ref#1{Ref.~\cite{#1}}
\def\Refs#1{Refs.~\cite{#1}}

%use footnote style for a bibitem ref in latex
\def\Cite#1{${\mathstrut}^{\cite{#1}}$}

% abbrevs for latex commands:
%equations
\def\EQ{\begin{equation}}
\def\endEQ{\end{equation}}

%my macros

\def\fewquad{\qquad\qquad}
\def\severalquad{\qquad\fewquad}
\def\manyquad{\qquad\severalquad}
\def\manymanyquad{\manyquad\manyquad}

\def\downindex#1{{}_{#1}}
\def\upindex#1{{}^{#1}}

\def\mathtext#1{\hbox{\rm{#1}}}

\def\hp#1{\hphantom{#1}}

\def\parder#1#2{\partial{#1}/\partial{#2}}
\def\parderop#1{\partial/\partial{#1}}

\def\sech{{\rm sech}}
\def\sgn{{\rm sgn}}

\def\ie/{i.e.}
\def\eg/{e.g.}
\def\etc/{etc.}
\def\const{{\rm const.}}

%\begin{frontmatter}

\title{Interaction properties of complex mKdV solitons}

\author{Stephen C. Anco}
\address{
Department of Mathematics,
Brock University,
St. Catharines, ON L2S 3A1 Canada }

\author{Nestor Tchegoum Ngatat}
\address{
Department of Mathematics,
Brock University,
St. Catharines, ON L2S 3A1 Canada }

\author{Mark Willoughby}
\address{
Institute for Applied Mathematics,
University of British Columbia,
Vancouver, BC Canada }

\begin{abstract}
Interaction properties of complex solitons are studied for 
the two $U(1)$-invariant integrable generalizations of the mKdV equation, 
given by the Hirota equation and the Sasa-Satsuma equation, 
which share the same travelling wave (single-soliton) solution 
having a {\em sech} profile characterized by 
a constant speed 
and a constant phase angle. 
%$c>0$ $-\pi\leq\phi\leq\pi$
For both equations, 
nonlinear interactions where a fast soliton 
%with speed $c_1$ and phase $\phi_1$
collides with a slow soliton 
%with speed $c_2$ and phase $\phi_2$, 
are shown to be described by $2$-soliton solutions that can have
three different types of interaction profiles 
depending on the speed ratio 
%$r=c_1/c_2$ 
and the relative phase angle 
%$\Delta\phi=\phi_1-\phi_2$
of the individual solitons. 
In all cases the shapes and speeds of the solitons 
are found to be preserved apart from a shift in position such that 
their center of momentum moves at a constant speed. 
Moreover, for the Hirota equation, 
the phase angles of the fast and slow solitons are found to remain unchanged,
while for the Sasa-Satsuma equation, 
the phase angles are shown to undergo a shift such that 
the relative phase between the fast and slow solitons changes sign. 
\end{abstract}

%\begin{keyword}
%\MSC
%\end{keyword}
\def\sep{;}
\keywords{
mKdV, solitary wave, soliton, collision, interaction profile, position shift, 
phase shift
}
\subjclass[2010]{}
%\thanks{}

\maketitle

\section{ Introduction }

Solitons are solitary waves that retain their shape and speed 
after undergoing collisions or other nonlinear interactions
\cite{AblCla}. 
A widely familiar example is the {\em sech}{}$^2$
solitary wave solution of the Korteweg-de Vries (KdV) equation
$u_t +6u u_x +u_{xxx} =0$. 
This solution 
$u(t,x) = (c/2)\sech^2\left(\sqrt{c}(x-ct)/2\right)$
is a stable travelling wave that is single-peaked 
and uni-directional. 
It carries mass, momentum, energy, as well as Galilean momentum 
(associated with the motion of center of mass), 
which are constants of motion for the KdV equation. 
A collision occurs when a faster solitary wave overtakes 
a slower solitary wave. 
(See the animations at 
http://lie.math.brocku.ca/\url{~}sanco/solitons/kdv\_solitons.php
)
Remarkably, the only net effect of the collision is that 
the faster wave is shifted forward in position
while the slower wave is shifted backward in position,
where these shifts depend solely on the speeds of both waves
and do not affect the center of mass of the two waves 
(which moves at constant speed throughout the collision). 

During a collision, 
KdV solitary waves interact nonlinearly such that \cite{Lax,LeV}
their peaks either first merge together and then split apart 
if the speed ratio of the waves is greater than $3$
or first bounce and then exchange shapes and speeds
if the speed ratio of the waves is less than $3$. 
There are many different alternative ways to interpret this nonlinear interaction, 
such as the faster wave getting stretched 
while the slower wave is squeezed underneath it, 
or as the faster wave emitting an intermediate wave that is absorbed
by slower wave.
(See \Ref{BenKasYou} for a comprehensive survey of interpretations.) 
The same interaction properties also hold more generally 
for pair-wise collisions of any number of KdV solitary waves. 

Solutions of the KdV equation describing 
collisions of $n>1$ solitary waves are called $n$-solitons 
and have been obtained by many different methods 
(\eg/ auto-Backlund transformations, nonlinear superposition formula, 
Hirota bilinear equations, inverse scattering, dressing equations),
all of which rely on the underlying integrability of the KdV equation, 
in particular the existence of a Lax pair \cite{Lax}. 
An important consequence of this integrability is that 
KdV solitons have constants of motion consisting of 
mass, momentum, Galilean momentum, and energy, 
which are defined by conserved integrals \cite{Whibook} 
involving $t,x,u,u_x$, 
plus an infinite number of higher-order ``energies'' 
(involving higher order $x$-derivatives of $u$) \cite{AblCla}. 

In this paper, 
we study the collision properties of solitons of 
the modified KdV (mKdV) equation
\EQ
u_t +\alpha u^2 u_x +\beta u_{xxx} =0
\label{mkdveq}
\endEQ
and its integrable $U(1)$-invariant generalizations
\begin{align}
u_t +\alpha |u|^2 u_x +\beta u_{xxx} =0
\label{hmkdveq}\\
u_t +\tfrac{1}{4}\alpha( u\bar u_x +3u_x\bar u )u +\beta u_{xxx} =0
\label{ssmkdveq}
\end{align}
where $\bar u$ denotes the complex conjugate of $u$,
and $|u|$ denotes the modulus of $u$. 
Here $\alpha$ and $\beta$ are arbitrary positive constants. 
These two generalizations are known to be 
\cite{complexmkdveqs1,complexmkdveqs2}
the only complex versions of the mKdV equation 
that possess a Lax pair with the same scaling symmetry
\EQ
t\rightarrow \lambda^3 t,\quad
x\rightarrow \lambda x,\quad
u\rightarrow \lambda^{-1} u 
\label{scalingsymm}
\endEQ
admitted by the mKdV equation \eqref{mkdveq}. 
Both generalizations also have an additional $U(1)$ phase symmetry 
\EQ
u\rightarrow \exp(i\phi) u 
\label{phasesymm}
\endEQ
In this form, 
all three equations \eqref{mkdveq}, \eqref{hmkdveq}, \eqref{ssmkdveq}
share the same solitary wave solution (\ie/ a $1$-soliton)
\EQ
u(t,x) = \pm\sqrt{\frac{6c}{\alpha}} \sech\left(\sqrt{\frac{c}{\beta}}(x-ct)\right)
\label{realsoliton}
\endEQ
where $c>0$ is the wave speed. 
As shown by the results in \Ref{conslaws}
on constants of motion for equations of complex mKdV form, 
the solitary wave solution \eqref{realsoliton} has mass, momentum and energy, 
which are given by counterparts of the KdV conserved integrals 
involving just $u,u_x$; 
in addition, 
although there is no counterpart of KdV Galilean momentum for this solution, 
it has an analogous Galilean energy given by a conserved integral
which is related to the motion of center of momentum. 

In the case of the real mKdV equation \eqref{mkdveq}, 
the $1$-soliton \eqref{realsoliton} has an up or down orientation 
corresponding to the plus or minus sign of $u$, 
which comes from the discrete reflection symmetry $u\rightarrow -u$ 
of this equation. 
Thus, 
there are two different types of real mKdV soliton collisions,
where (up to reflection) the fast and slow solitons in the collision
have either the same orientation or opposite orientations. 

In both cases of the complex mKdV equations \eqref{hmkdveq} and \eqref{ssmkdveq}, 
the sign of the $1$-soliton \eqref{realsoliton} can be absorbed 
into an arbitrary constant phase
\EQ
u(t,x) = \sqrt{\frac{6c}{\alpha}} \exp(i\phi) \sech\left(\sqrt{\frac{c}{\beta}}(x-ct)\right) ,\quad
\phi=\const
\label{1soliton}
\endEQ
due to the $U(1)$ symmetry \eqref{phasesymm}.
Consequently, 
collisions of two complex mKdV solitons \eqref{1soliton} 
involve a relative phase angle, given by the difference of 
the phase angles of the fast and slow solitons in the collision. 
This relative phase therefore 
parameterizes the types of collisions. 
An interesting question we will study is whether the two phase angles
are altered in a collision, 
\ie/ do the fast and slow solitons each get shifted in phase 
as well as in position?

In \secref{mkdvcollision}, 
the $2$-soliton solution that describes collisions of 
fast and slow solitary waves for the mKdV equation \eqref{mkdveq} 
is reviewed. 
The properties of these collisions depend only on 
the ratio of speeds and the relative orientation of the two waves. 
In particular, we show that 
if the waves have the same orientation 
then their nonlinear interaction in a collision 
either is a merge-split type when their speed ratio is greater than 
the value $(7+3\sqrt{5})/2$,
or otherwise is a bounce-exchange type when their speed ratio is less than 
this critical value $(7+3\sqrt{5})/2$. 
In contrast, 
if the waves have opposite orientations 
then their nonlinear interaction instead is a completely different type 
in which (regardless of their speed ratio) 
the slow soliton gradually is first absorbed 
and then emitted by the fast soliton. 
In all three interactions, 
we show that the net effect of the collision is solely that 
the faster wave is shifted forward in position
while the slower wave is shifted backward in position,
such that the center of momentum of the two waves is unaffected
(moving at a constant speed throughout the collision). 

In \secref{hmkdvcollision}, 
we consider the complex mKdV equation \eqref{hmkdveq},
which is commonly called the Hirota-mKdV equation
\cite{Hir1973}. 
We carry out an asymptotic analysis of the $2$-soliton solution 
describing collisions of fast and slow solitary waves. 
Our results show that the interaction properties of the two waves 
depend only on their speed ratio and their relative phase angle
such that a bounce-exchange type of interaction occurs 
when the speed ratio is less than a critical value given by a 
certain explicit function of the relative phase angle
and that otherwise a merge-split or absorb-emit type of interaction occurs
depending on whether the relative phase angle is less than or greater than 
a certain critical value in terms of the speed ratio of the waves. 
For each type of interaction, 
we find that the phase angles of the two waves remain unchanged 
and the only effect of the collision is to produce 
a respective forward and backward shift in the positions 
of the faster and slower waves. 
In particular, 
these shifts are found to depend only on the speeds of the two waves, 
but not on their phases angles, 
such that the center of momentum of the two waves moves at a constant speed 
throughout the collision. 

In \secref{ssmkdvcollision},
we consider the other complex mKdV equation \eqref{ssmkdveq},
which is known as the Sasa-Satsuma-mKdV equation
\cite{SasSat1991}. 
We write down the $2$-soliton solution in an explicit form 
parameterized by the speeds and phase angles of the fast and slow solitary waves
(which has not appeared previously in the literature). 
Through an asymptotic analysis of this solution, 
we find the interesting new result that the phase angles of the two waves
in a collision undergo a shift such that the relative phase angle changes sign. 
In addition, the positions of the fast and slow waves display 
a respective forward and backward shift which depends on both 
the speeds and the relative phase angle of the waves. 
We show that these position shifts preserve the center of momentum 
of the two waves in the collision, 
while the phase-angle shifts are related to an invariance property of 
$2$-soliton solution with respect to space-time reflection 
combined with phase conjugation. 
Finally, 
we also derive the detailed interaction properties of the two waves.
We show that the waves exhibit 
a bounce-exchange type of interaction only when 
their the relative phase angle is less than $\arccos(-1/6) \approx 0.55\pi$ 
and their speed ratio is less than a critical value given by a 
certain explicit function of the relative phase angle 
(which is different than the function arising for the Hirota-mKdV $2$-soliton solution). 
For any speed ratio greater than this value, 
or for any relative phase angle greater than $\arccos(-1/6) \approx 0.55\pi$,
we find that the waves exhibit a merge-split type of interaction 
if their relative phase angle is less than a certain critical value 
in terms of the their speed ratio 
(which is again different than the critical angle found for the Hirota-mKdV $2$-soliton solution), 
and that otherwise the waves exhibit an absorb-emit type of interaction. 

Last, 
some features of the soliton collisions for the Hirota and Sasa-Satsuma equations 
are compared in section \secref{compare}. 

In the appendix, 
we provide a short derivation of the $2$-soliton solution 
for the Sasa-Satsuma-mKdV equation. 
Hereafter, by scaling variables, we will put 
\EQ
\alpha=24, \quad
\beta=1
\endEQ 
for convenience.

\section{real mKdV soliton collisions}
\label{mkdvcollision} 
For the mKdV equation
\EQ
u_t +24 u^2 u_x + u_{xxx} =0
\label{mkdv}
\endEQ
we first recall the conserved integrals defining counterparts of
KdV mass, momentum, and energy. 
These integrals are given by \cite{MiuGarKru}
\begin{align}
& \mathcal{M}= \int_{-\infty}^{+\infty} u\; dx
\label{m}\\
& \mathcal{P}= \int_{-\infty}^{+\infty} u^2\; dx
\label{p}\\
& \mathcal{E}= \int_{-\infty}^{+\infty} \frac{1}{2}u_x^2 -2u^4\; dx
\label{e}
\end{align}
Although there is no counterpart of KdV Galilean momentum, 
the mKdV equation has the extra conserved integral \cite{MiuGarKru}
\EQ
\mathcal{C}= \int_{-\infty}^{+\infty} t(\frac{1}{2}u_x^2 -2u^4)-x\frac{1}{6}u^2\; dx
\label{Gale}
\endEQ
which defines a Galilean energy related to center of momentum 
as given by 
\EQ
\mathcal{X}(t)= \frac{1}{\mathcal{P}} \int_{-\infty}^{+\infty} x u^2\; dx
= \mathcal{X}(0) +6\frac{\mathcal{E}}{\mathcal{P}} t
\label{com}
\endEQ
where
\EQ
\mathcal{C}= t\mathcal{E} -\frac{1}{6}\mathcal{P}\mathcal{X}(t) 
= \mathcal{C}(0)= -\frac{1}{6}\mathcal{P}\mathcal{X}(0) 
\endEQ
These integrals \eqref{m}--\eqref{Gale} are constants of motion 
for all smooth solutions $u(t,x)$ with asymptotic decay $u\rightarrow 0$
as $x\rightarrow\pm\infty$. 
In particular, this includes the solutions describing 
solitary waves and their collisions. 

The $1$-soliton for the mKdV equation is given by \cite{Hir1972}
\EQ
u(t,x) = \pm\frac{\sqrt{c}}{2} \sech\left(\sqrt{c}\xi\right)
= s\frac{\sqrt{c} \exp(\sqrt{c}\xi)}{1+\exp(2\sqrt{c}\xi)} 
\label{mkdv1soliton}
\endEQ
with speed $c>0$ and up/down orientation $s=\pm 1$, 
where
\EQ
\xi = x-ct
\label{movingcoord}
\endEQ
is a moving coordinate. 
This solution describes a stable travelling wave that is single-peaked 
and uni-directional. 
Its height relative to $u=0$ is $\pm\sqrt{c}/2$, 
and its width is proportional to $1/\sqrt{c}$. 
It has constants of motion 
\EQ
\mathcal{M}=s\frac{\pi}{2} ,\quad
\mathcal{P}=\frac{\sqrt{c}}{2} ,\quad
\mathcal{E}=\frac{\sqrt{c}^3}{12} ,\quad
\mathcal{C}=0 , 
\endEQ
while its center of momentum is 
\EQ
\mathcal{X}(t)= 6\frac{\mathcal{E}}{\mathcal{P}} t = ct
\endEQ
which coincides with the position of the peak. 
The initial position of the wave can be shifted arbitrarily by means of 
a space translation $x\rightarrow x-x_0$ 
applied to the moving coordinate \eqref{movingcoord}, 
so then $\xi=x-ct-x_0$ and $\mathcal{X}(t)= x_0+ ct$. 
This changes the Galilean energy of the wave to be  
\EQ
\mathcal{C}= -x_0\sqrt{c}/12 , 
\endEQ
while the mass, momentum and energy are unchanged.

\subsection{$2$-soliton solution}
\noindent\newline\indent
Collisions where a fast soliton with speed $c_1$ 
overtakes a slow soliton with speed $c_2$
are described by the well-known $2$-soliton solution \cite{Hir1972}
which depends on the orientations $s_1$ and $s_2$ 
of the respective solitons. 
(See the animations of collisions at 
http://lie.math.brocku.ca/\url{~}sanco/solitons/mkdv\_solitons.php
)
We will write this solution in the rational-exponential form 
\EQ
u(t,x) = \frac{G}{F}
\label{mkdv2soliton}
\endEQ
given by 
\begin{align}
G =& \kappa \big(
s_1 \sqrt{c_1}\exp(\sqrt{c_1}\xi_1)(1+\exp(2\sqrt{c_2}\xi_2))
+ s_2 \sqrt{c_2}\exp(\sqrt{c_2}\xi_2)(1+\exp(2\sqrt{c_1}\xi_1)) 
\big) 
\label{mkdvG}\\
F=& 1+ 2 s_1 s_2 (\kappa^2-1) \exp(\sqrt{c_1}\xi_1 +\sqrt{c_2}\xi_2) 
+\kappa^2( \exp(2\sqrt{c_1}\xi_1) + \exp(2\sqrt{c_2}\xi_2) )
\nonumber\\&\qquad
+ \exp(2(\sqrt{c_1}\xi_1 +\sqrt{c_2}\xi_2))
\label{mkdvF}
\end{align}
with 
\EQ
s_1=\pm 1 ,\quad
s_2=\pm 1 ,\quad
\kappa =\frac{\sqrt{c_1}+\sqrt{c_2}}{\sqrt{c_1}-\sqrt{c_2}} >1
\endEQ
where
\EQ
\xi_1 = x-c_1 t-x_1, \quad
\xi_2 = x-c_2 t-x_2
\label{movingcoords}
\endEQ
are moving coordinates 
centered at initial positions $x=x_1$ and $x=x_2$ respectively. 

We now examine the asymptotic form of 
the $2$-soliton solution \eqref{mkdv2soliton}--\eqref{mkdvF} 
as $t\rightarrow\pm\infty$. 

To proceed, 
we consider $\xi_1=\const$ and express $\xi_2=\xi_1+\zeta$
in terms of $\xi_1$ and $\zeta=t\Delta c-\Delta x$,
where
$\Delta c=c_1-c_2>0$ is the relative speed of the moving coordinates,
and 
$\Delta x=x_2-x_1$ is the separation of the centers of the moving coordinates. 
Note $t\rightarrow\pm\infty$ corresponds to $\zeta\rightarrow\pm\infty$. 
We then asymptotically expand $F$ and $G$ for large $\zeta$
with $\xi_1$ held fixed.
This yields, after neglecting subdominant exponential terms, 
\begin{align}
& G\simeq
\begin{cases}
s_1 \kappa\sqrt{c_1} \exp((\sqrt{c_1} +2\sqrt{c_2})\xi_1) 
\exp(2\sqrt{c_2}\zeta) 
& \zeta\rightarrow +\infty\\
s_1 \kappa\sqrt{c_1} \exp(\sqrt{c_1}\xi_1)
& \zeta\rightarrow -\infty
\end{cases}
\\
& F\simeq
\begin{cases}
\big( \kappa^2 \exp(2\sqrt{c_2}\xi_1) + \exp(2(\sqrt{c_1} +\sqrt{c_2})\xi_1) \big)
\exp(2\sqrt{c_2}\zeta) 
& \zeta\rightarrow +\infty\\
1+\kappa^2 \exp(2\sqrt{c_1}\xi_1) 
& \zeta\rightarrow -\infty
\end{cases}
\end{align}
and hence
\begin{align}
& u\simeq 
\begin{cases}
s_1\dfrac{(\sqrt{c_1}/\kappa) \exp(\sqrt{c_1}\xi_1)}
{1+(1/\kappa^2)\exp(2\sqrt{c_1}\xi_1)} 
& \zeta\rightarrow +\infty\\
s_1\dfrac{\sqrt{c_1} \kappa \exp(\sqrt{c_1}\xi_1)}
{1+\kappa^2 \exp(2\sqrt{c_1}\xi_1)} 
& \zeta\rightarrow -\infty
\end{cases}
\end{align}
Thus, 
in this expansion the $2$-soliton solution asymptotically reduces to 
the form of a $1$-soliton solution 
\EQ
u(t,x) \simeq 
s_1 \frac{\sqrt{c_1} \exp(\sqrt{c_1}\xi_1^\pm)}{1+\exp(2\sqrt{c_1}\xi_1^\pm)} 
=u_1^\pm ,\quad
t\rightarrow\pm\infty ,\quad
\xi_1=\const
\label{mkdvfastsoliton}
\endEQ
in terms of a moving coordinate 
\EQ
\xi_1^\pm = \xi_1 -a_1^\pm 
\label{shiftfastcoord}
\endEQ
which is shifted relative to $\xi_1$ by 
\EQ
a_1^\pm = \pm\ln(\kappa)/\sqrt{c_1}
\label{a1}
\endEQ

To continue, 
we next consider $\xi_2=\const$ and express $\xi_1=\xi_2-\zeta$
in terms of $\xi_2$ and $\zeta=t\Delta c-\Delta x$ again. 
By asymptotically expanding $F$ and $G$ for large $\zeta$
with $\xi_2$ held fixed, 
and neglecting subdominant exponential terms, 
we obtain 
\begin{align}
& G\simeq
\begin{cases}
s_2\kappa\sqrt{c_2} \exp(\sqrt{c_2}\xi_2)
& \zeta\rightarrow +\infty\\
s_2\kappa\sqrt{c_2} \exp((\sqrt{c_2} +2\sqrt{c_1})\xi_2) 
\exp(-2\sqrt{c_1}\zeta) 
& \zeta\rightarrow -\infty
\end{cases}
\\
& F\simeq
\begin{cases}
1+\kappa^2 \exp(2\sqrt{c_2}\xi_2) 
& \zeta\rightarrow +\infty\\
\big( \kappa^2 \exp(2\sqrt{c_1}\xi_2) + \exp(2(\sqrt{c_2} +\sqrt{c_1})\xi_2) \big)
\exp(-2\sqrt{c_1}\zeta) 
& \zeta\rightarrow -\infty
\end{cases}
\end{align}
and thus 
\begin{align}
& u\simeq 
\begin{cases}
s_2\dfrac{\sqrt{c_2} \kappa \exp(\sqrt{c_2}\xi_2)}
{1+\kappa^2 \exp(2\sqrt{c_2}\xi_2)} 
& \zeta\rightarrow +\infty\\
s_2\dfrac{(\sqrt{c_2}/\kappa) \exp(\sqrt{c_2}\xi_2)}
{1+(1/\kappa^2)\exp(2\sqrt{c_2}\xi_2)} 
& \zeta\rightarrow -\infty
\end{cases}
\end{align}
This asymptotic expansion of the $2$-soliton solution 
again has the form of a $1$-soliton solution 
\EQ
u(t,x) \simeq 
s_2 \frac{\sqrt{c_2} \exp(\sqrt{c_2}\xi_2^\pm)}{1+\exp(2\sqrt{c_2}\xi_2^\pm)} 
=u_2^\pm ,\quad
t\rightarrow\pm\infty ,\quad
\xi_2=\const
\label{mkdvslowsoliton}
\endEQ
in terms of a moving coordinate 
\EQ
\xi_2^\pm = \xi_2 -a_2^\pm 
\label{shiftslowcoord}
\endEQ
which is shifted relative to $\xi_2$ by 
\EQ
a_2^\pm = \mp\ln(\kappa)/\sqrt{c_2}
\label{a2}
\endEQ

\subsection{Constants of motion and asymptotic position shifts}
\begin{figure}[!h]
\centering\includegraphics[scale=0.85]{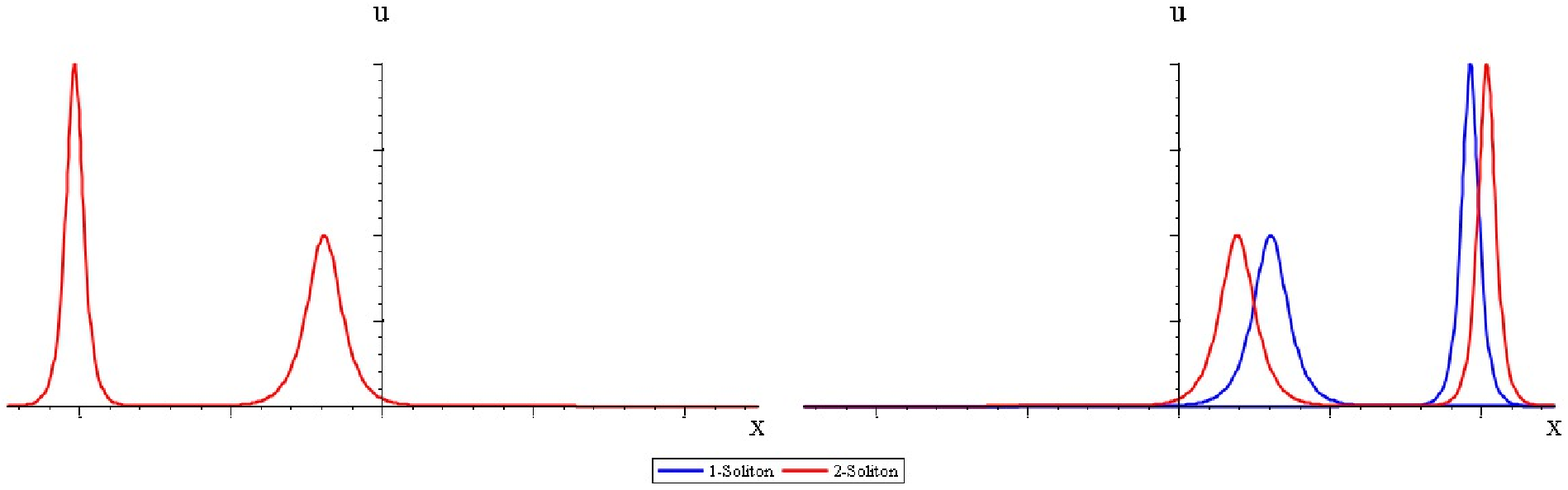}
\caption{mKdV equation $2$-soliton interaction with 
$c_1/c_2=2$, $s_1/s_2=1$}
\label{mkdv-shift-same}
\end{figure}

\begin{figure}[!h]
\centering\includegraphics[scale=0.85]{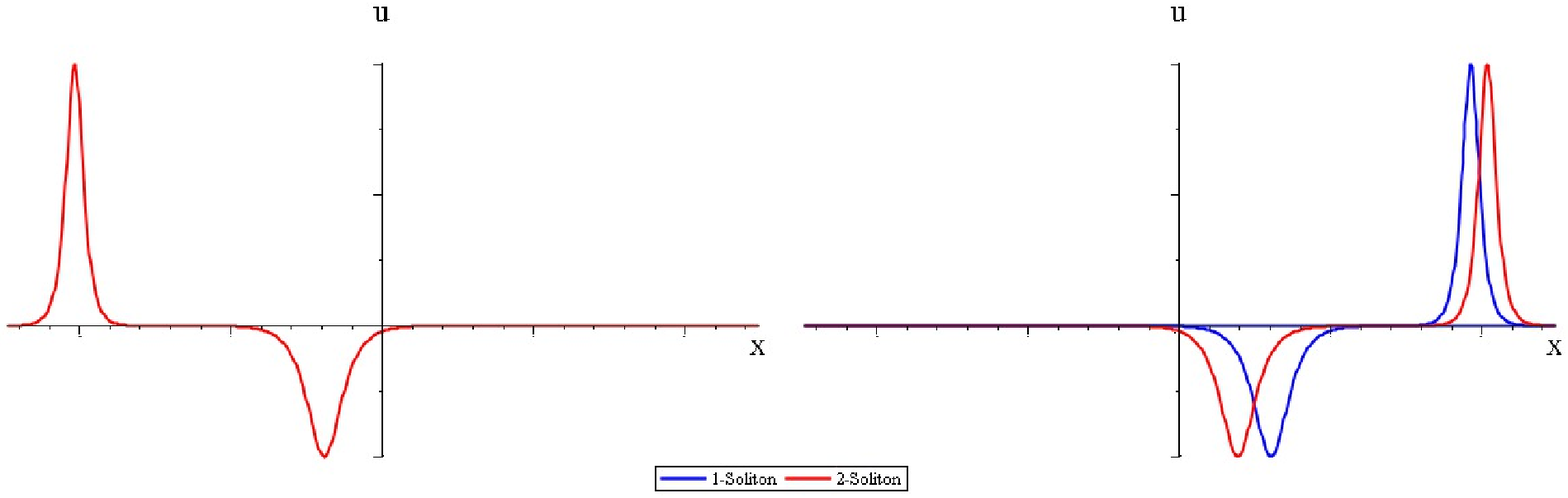}
\caption{mKdV equation $2$-soliton interaction with 
$c_1/c_2=2$, $s_1/s_2=-1$}
\label{mkdv-shift-opposite}
\end{figure}

\noindent\newline\indent
The expansions \eqref{mkdvfastsoliton} and \eqref{mkdvslowsoliton}
show that for $t\rightarrow \pm\infty$
the $2$-soliton solution \eqref{mkdv2soliton}--\eqref{mkdvF} 
asymptotically has the form of a superposition $u\simeq u_1^\pm+u_2^\pm$ 
of a fast soliton $u_1^\pm$ with speed $c_1$ and up/down orientation $s_1$ 
and a slow soliton $u_2^\pm$ with speed $c_2$ and up/down orientation $s_2$,
whose positions are determined by the moving coordinates 
$\xi_1^\pm$ and $\xi_2^\pm$. 
(This result generalizes the well-known analysis in \Ref{Wad}
which considered the case $s_1=s_2=1$.)
See \figref{mkdv-shift-same} and \figref{mkdv-shift-opposite}. 

In the asymptotic past ($t\rightarrow-\infty$), 
$\xi_1^-=0$ gives the position of the peak of the fast soliton, 
while in the asymptotic future ($t\rightarrow+\infty$), 
the position of the peak is instead given by $\xi_1^+=0$. 
These asymptotic positions lie on straight lines in space-time
\EQ
x = x_1 +a_1^\mp +c_1 t
\endEQ
Comparing the asymptotic past with the asymptotic future,
we see that the fast soliton retains its shape and speed
but gets shifted forward in position as given by 
\EQ
\Delta x_1 = a_1^+ - a_1^- 
=\frac{2}{\sqrt{c_1}} \ln\left(\frac{\sqrt{c_1}+\sqrt{c_2}}{\sqrt{c_1}-\sqrt{c_2}}\right)
% 2\ln(\kappa)/\sqrt{c_1} 
>0
\label{mkdvfastshift}
\endEQ

Similarly, the slow soliton retains its shape and speed,
while the position of its peak 
in the asymptotic past and future lies on the straight lines
$x=x_2 +a_2^\mp +c_2 t$. 
So we see that the slow soliton 
gets shifted backward in position as given by 
\EQ
\Delta x_2 = a_2^+ - a_2^- 
= -\frac{2}{\sqrt{c_2}} \ln\left(\frac{\sqrt{c_1}+\sqrt{c_2}}{\sqrt{c_1}-\sqrt{c_2}}\right)
%-2\ln(\kappa)/\sqrt{c_2} 
<0
\label{mkdvslowshift}
\endEQ

The asymptotic shifts \eqref{mkdvfastshift} and \eqref{mkdvslowshift}
do not depend on the orientations of the fast and slow solitons. 
Moreover, these shifts satisfy the relation 
\EQ
\sqrt{c_1}\Delta x_1 + \sqrt{c_2}\Delta x_2 =0
\label{mkdvshifteq}
\endEQ
which can be understood as a consequence of the motion of 
the center of momentum of the $2$-soliton solution 
(similarly to the same result known for the KdV 2-soliton solution
\cite{WadTod}).

In particular, 
because $u\simeq u_1^\pm+u_2^\pm$ is a superposition as $t\rightarrow\pm\infty$,
the conserved mass, momentum and energy of $u$ are given by 
\begin{align}
&\mathcal{M}=\mathcal{M}_1 +\mathcal{M}_2
=(s_1+s_2)\frac{\pi}{2} 
\\
&\mathcal{P}=\mathcal{P}_1 +\mathcal{P}_2
= \frac{\sqrt{c_1}+\sqrt{c_2}}{2} 
\label{mkdv2solitonp}\\
&\mathcal{E}=\mathcal{E}_1 +\mathcal{E}_2 
=\frac{\sqrt{c_1}^3 + \sqrt{c_2}^3}{12} 
\label{mkdv2solitone}
\end{align}
in terms of the mass, momentum and energy individually associated with
the fast and slow solitons $u_1^\pm$ and $u_2^\pm$.
Similarly, 
the conserved Galilean energy of $u$ is given by 
\EQ
\mathcal{C}=\mathcal{C}_1 + \mathcal{C}_2
= -\frac{(x_1+a_1^-) \sqrt{c_1}+ (x_2+a_2^-) \sqrt{c_2}}{12}
= -\frac{x_1+a_1^+) \sqrt{c_1}+ (x_2 +a_2^+) \sqrt{c_2}}{12}
\endEQ
which simplifies to 
\EQ
\mathcal{C}= -\frac{x_1 \sqrt{c_1}+ x_2 \sqrt{c_2}}{12}
= t\mathcal{E}-\frac{1}{6}\mathcal{P}\mathcal{X}(t)
\label{mkdv2solitongale}
\endEQ
due to the relation $a_1^\pm\sqrt{c_1} + a_2^\pm\sqrt{c_2}=0$. 
Thus the center of momentum of $u$ is 
\EQ
\mathcal{X}(t)
= \frac{\sqrt{c_1}(x_1+c_1 t) + \sqrt{c_2}(x_2+c_2 t)}{\sqrt{c_1}+ \sqrt{c_2}}
= \frac{\mathcal{P}_1\mathcal{X}_1^\pm(t) + \mathcal{P}_2\mathcal{X}_2^\pm(t)}{
\mathcal{P}_1 + \mathcal{P}_2}
\label{mkdv2solitoncom}
\endEQ
where 
$\mathcal{X}_1^\pm(t)=x_1+a_1^\pm +c_1 t$ 
and $\mathcal{X}_2^\pm(t)=x_2 +a_2^\pm +c_2 t$ 
are the respective centers of momentum of the fast and slow solitons
in the asymptotic past and future. 
Therefore, we see that the center of momentum of the $2$-soliton solution 
moves at constant speed
\EQ
c=\frac{\sqrt{c_1}^3 + \sqrt{c_2}^3}{\sqrt{c_1} + \sqrt{c_2}}
= \frac{\mathcal{P}_1 c_1 + \mathcal{P}_2 c_2}{\mathcal{P}_1 + \mathcal{P}_2}
\label{mkdv2solitonspeed}
\endEQ
and consequently the asymptotic shifts $\Delta x_1$ and $\Delta x_2$ 
in the positions of the two solitons $u_1^\pm$ and $u_2^\pm$ are constrained to satisfy
$\mathcal{P}_1 \Delta x_1 + \mathcal{P}_2 \Delta x_2 =0$
which explains the relation \eqref{mkdvshifteq}.

\subsection{Interaction profile}
\noindent\newline\indent
We now study the interaction profile of 
the $2$-soliton solution \eqref{mkdv2soliton}--\eqref{mkdvF},
which we will write in the equivalent form 
\EQ
u(t,x) = 
\frac{\kappa( s_1 \sqrt{c_1}\cosh(\theta_2) + s_2 \sqrt{c_2}\cosh(\theta_1) )}
{s_1 s_2 (\kappa^2-1) +\kappa^2 \cosh(\theta_1-\theta_2) + \cosh(\theta_1 +\theta_2)}
\label{mkdv2soliton'}
\endEQ
in terms of 
\EQ
\theta_1 = \sqrt{c_1}\xi_1 ,\quad
\theta_2 = \sqrt{c_2}\xi_2 
\endEQ
For simplicity, by means of suitable time and space translations
$t\rightarrow t-t_0$, $x\rightarrow x-x_0$, 
we shift the centers of the moving coordinates \eqref{movingcoords}
to the positions
\EQ
x_1=x_0-c_1 t_0 =0,\quad
x_2=x_0-c_2 t_0 =0
\endEQ
Then the resulting $2$-soliton solution \eqref{mkdv2soliton'} 
is invariant under a combined space-time reflection 
$x\rightarrow -x$, $t\rightarrow -t$,
and its center of momentum is simply $\mathcal{X}(t)=ct$
in terms of the speed \eqref{mkdv2solitonspeed}. 
In the asymptotic past and future ($t\rightarrow \mp\infty$), 
this solution describes a superposition 
$u\simeq u_1^\pm +u_2^\pm$ of fast and slow solitons $u_1^\pm$ and $u_2^\pm$
whose centers of momentum are given by 
$\mathcal{X}_1^\mp(t)=a_1^\mp +c_1 t$ and $\mathcal{X}_2^\mp(t)=a_2^\mp +c_2 t$
where
\EQ
\Delta \mathcal{X}_1^\mp(t)=\mathcal{X}_2^\mp(t)- \mathcal{X}_1^\mp(t)
= \pm \frac{1}{2}(\Delta x_1 -\Delta x_2) -(c_1-c_2) t
\label{mkdvu1u2separation}
\endEQ
is the separation between the peaks of $u_1^\pm$ and $u_2^\pm$
in terms of the asymptotic shifts \eqref{mkdvfastshift} and \eqref{mkdvslowshift} for $t\rightarrow \mp\infty$.

Because of the space-time reflection invariance of $u$,
the separation between the fast and slow peaks in $u$ 
will be a minimum at time $t=0$ when $u$ is an even function of $x$. 
Qualitatively speaking, 
$t=0$ will be the moment of greatest nonlinear interaction 
between the fast and slow solitons. 
We will therefore refer to the shape of $u$ at $t=0$ 
as the {\em interaction profile} of the $2$-soliton solution. 
Since this profile $u(0,x)$ is even in $x$, 
its shape can be characterized by the convexity 
\EQ
u(0,x)_{xx}\big|_{x=0} =\frac{1}{2}(c_1-c_2)(\sqrt{c_1} -\sqrt{c_2})
( \sqrt{c_1 c_2}-(s_1\sqrt{c_1} -s_2\sqrt{c_2})^2 )
\label{mkdv2solitonconvexity}
\endEQ
Note that the sign of the convexity \eqref{mkdv2solitonconvexity} 
depends only on the ratio of speeds and the relative orientation 
\EQ
r=c_1/c_2 ,\quad
s=s_1/s_2 
\endEQ
of the fast and slow solitons. 
In particular, since $c_1>c_2>0$, we have
\EQ
\sigma= \sgn(u(0,x)_{xx}\big|_{x=0}) =\sgn( (2s+1)\sqrt{r}-r-1 ) 
\label{mkdvconvexitysgn}
\endEQ
with $r>1$ and $s=\pm 1$.

\begin{figure}[!h]
\centering\includegraphics[scale=0.85]{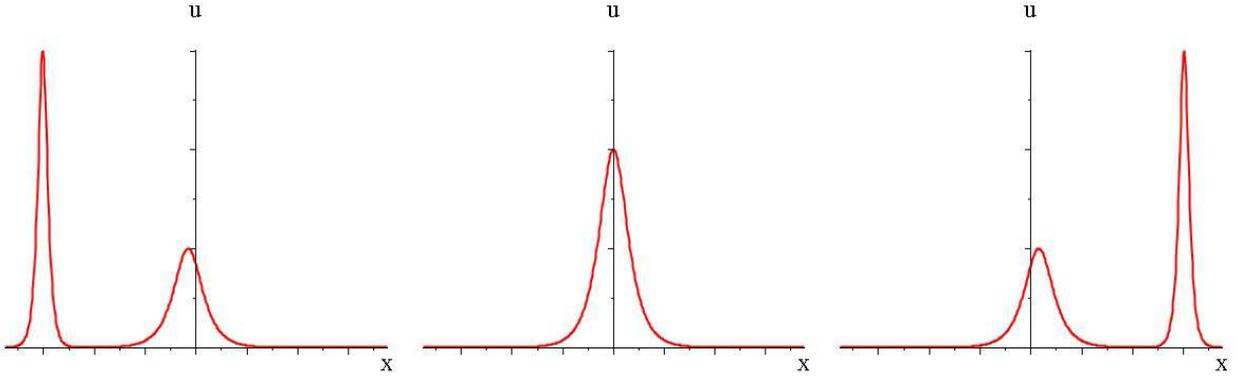}
\caption{mKdV equation $2$-soliton interaction with 
$c_1/c_2=9$, $s_1/s_2=1$}
\label{mkdv-same-merge}
\end{figure}

\begin{figure}[!h]
\centering\includegraphics[scale=0.85]{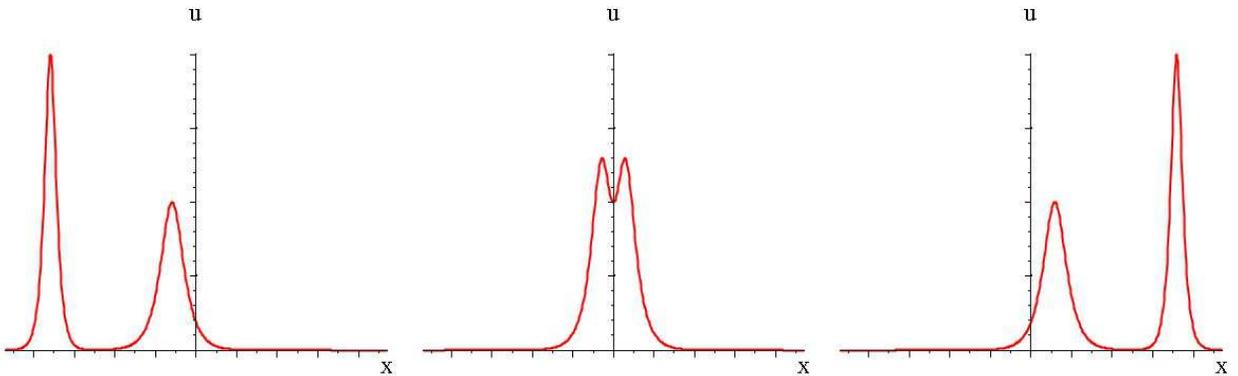}
\caption{mKdV equation $2$-soliton interaction with 
$c_1/c_2=4$, $s_1/s_2=1$}
\label{mkdv-same-bounce}
\end{figure}

The case where the two solitons have the same orientation is given by 
the relation $s=1$. 
In this case, the convexity sign is determined by 
\EQ
\sigma= \sgn(3\sqrt{r}-r-1) = \sgn( \tfrac{1}{2}(3+\sqrt{5})-\sqrt{r})
\endEQ
via factorization of \eqref{mkdvconvexitysgn} and use of the inequality $r>1$. 
Hence the sign is indefinite such that 
\begin{align}
\sigma 
\begin{cases}
<0 & \mathtext{ if } r> (7+3\sqrt{5})/2 \\
>0 & \mathtext{ if } r< (7+3\sqrt{5})/2 \\
=0 & \mathtext{ if } r= (7+3\sqrt{5})/2\\
\end{cases}
\label{mkdvcrit}
\end{align}
This result implies that the interaction profile $u(0,x)$ will have 
either a single peak at $x=0$ if $c_1/c_2 > (7+3\sqrt{5})/2$, 
or a double peak around $x=0$ if $c_1/c_2 < (7+3\sqrt{5})/2$.
These peaks will be positive or negative depending on the sign of $s_1$ ($=s_2$)
and will have an exponentially diminishing tail. 
In the case of a single peak,
the fast and slow solitons interact by 
first merging together at $t=x=0$ and then splitting apart,
while in the case of a double peak,
the fast and slow solitons interact by 
first bouncing and then exchanging shapes and speeds at $t=x=0$. 
We will call these cases, respectively, 
a {\em merge-split} and {\em bounce-exchange} interaction. 
See \figref{mkdv-same-merge} and \figref{mkdv-same-bounce}.

The value $c_1/c_2 = (7+3\sqrt{5})/2$ which separates 
these two types of interaction profiles
will be called the {\em critical speed ratio}. 

\begin{figure}[!h]
\centering\includegraphics[scale=0.85]{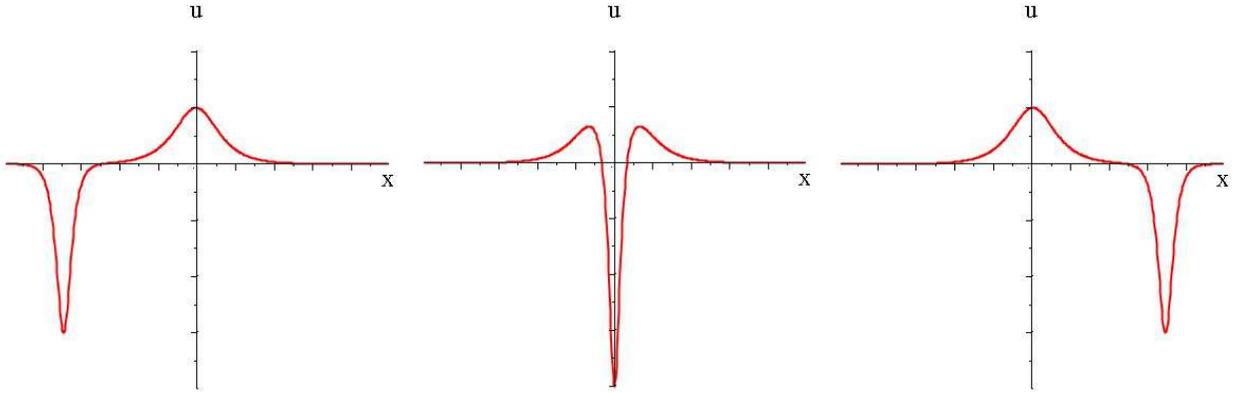}
\caption{mKdV equation $2$-soliton interaction with 
$c_1/c_2=9$, $s_1/s_2=-1$}
\label{mkdv-opposite}
\end{figure}

The other case, where the two solitons have opposite orientations, 
is given by $s=-1$.
In this case, 
the convexity sign \eqref{mkdvconvexitysgn} is strictly negative for all $r>1$,
\EQ
\sigma= -\sgn(\sqrt{r}+r+1)<0
\endEQ
implying that the interaction profile $u(0,x)$ will have 
a positive or negative peak at $x=0$ depending on the sign of $s_1$ ($=-s_2$).
Since $s_2$ has the opposite sign, the profile will also have
a pair of negative or positive side peaks around $x=0$, 
with an exponentially diminishing tail. 
The interaction between the fast and slow solitons in this case consists of 
the slow soliton gradually being first absorbed 
by the front side of the fast soliton
and then emitted from the back side of the fast soliton. 
We will call this an {\em absorb-emit} interaction. 
See \figref{mkdv-opposite}.

\section{Hirota-mKdV soliton collisions}
\label{hmkdvcollision} 
For the Hirota equation
\EQ
u_t +24 |u|^2 u_x + u_{xxx} =0
\label{h}
\endEQ
we note, firstly, 
there is no conserved integral for mass since 
neither $\int_{-\infty}^{+\infty} u\; dx$ nor $\int_{-\infty}^{+\infty} |u|\; dx$
is a constant of motion.
Secondly, 
we recall the conserved integrals for momentum, energy, and Galilean energy 
are given by \cite{conslaws}
\begin{align}
& \mathcal{P}= \int_{-\infty}^{+\infty} |u|^2\; dx
\label{cp}\\
& \mathcal{E}= \int_{-\infty}^{+\infty} \frac{1}{2}|u_x|^2 -2|u|^4\; dx
\label{ce}\\
& \mathcal{C}= \int_{-\infty}^{+\infty} t(\frac{1}{2}|u_x|^2 -2|u|^4)-x\frac{1}{6}|u|^2\; dx
\label{cGale}
\end{align}
which are constants of motion for 
all smooth solutions $u(t,x)$ with asymptotic decay $u\rightarrow 0$
as $x\rightarrow\pm\infty$. 
These integrals are related to the center of momentum 
\EQ
\mathcal{X}(t)= \frac{1}{\mathcal{P}} \int_{-\infty}^{+\infty} x |u|^2\; dx
= \mathcal{X}(0) +6\frac{\mathcal{E}}{\mathcal{P}} t
\label{ccom}
\endEQ
since
\EQ
\mathcal{C}= t\mathcal{E} -\frac{1}{6}\mathcal{P}\mathcal{X}(t) 
= \mathcal{C}(0)= -\frac{1}{6}\mathcal{P}\mathcal{X}(0) 
\label{ccomrel}
\endEQ
This is the same relation that holds for the mKdV constants of motion. 

The $1$-soliton solution for the Hirota equation is given by 
\EQ
u(t,x) = \frac{\sqrt{c}}{2}\exp(i\phi) \sech\left(\sqrt{c}\xi\right)
= \frac{\sqrt{c} \exp(i\phi+\sqrt{c}\xi)}{1+\exp(2\sqrt{c}\xi)} 
\label{cmkdv1soliton}
\endEQ
with speed $c>0$ and phase $-\pi\leq \phi\leq \pi$,
where
\EQ
\xi = x-ct
\label{movingcoord'}
\endEQ
is a moving coordinate. 
This solution describes a stable uni-directional travelling wave 
whose amplitude $|u|$ is the same as the amplitude of 
the mKdV solitary wave \eqref{mkdv1soliton}.
Therefore, 
its constants of motion \eqref{cp}--\eqref{cGale} and its center of momentum \eqref{ccom} 
are also the same as those for the mKdV solitary wave. 
In particular, 
the position of the peak amplitude coincides with the center of momentum
$\mathcal{X}(t)= ct$, which can be shifted arbitrarily by means of
a space translation 
\EQ
x\rightarrow x-x_0
\endEQ
applied to the moving coordinate \eqref{movingcoord'}.

\subsection{$2$-soliton solution}
\noindent\newline\indent
We now write down the $2$-soliton solution of the Hirota equation
describing collisions where a fast soliton with speed $c_1$ and phase $\phi_1$ 
overtakes a slow soliton with speed $c_2$ and phase $\phi_2$. 
(See the animations of collisions at 
http://lie.math.brocku.ca/\url{~}sanco/solitons/hirota.php
)
This solution has the rational-exponential form \eqref{mkdv2soliton}
given by \cite{Hir1973}
\begin{align}
G =& \kappa \big(
\sqrt{c_1}\exp(i\phi_1+\sqrt{c_1}\xi_1)(1+\exp(2\sqrt{c_2}\xi_2))
+ \sqrt{c_2}\exp(i\phi_2+ \sqrt{c_2}\xi_2)(1+\exp(2\sqrt{c_1}\xi_1)) 
\big) 
\label{hG}\\
F=& 1+ 2 \cos(\phi_1-\phi_2) (\kappa^2-1) \exp(\sqrt{c_1}\xi_1 +\sqrt{c_2}\xi_2) 
+\kappa^2( \exp(2\sqrt{c_1}\xi_1) + \exp(2\sqrt{c_2}\xi_2) )
\nonumber\\&\qquad
+ \exp(2(\sqrt{c_1}\xi_1 +\sqrt{c_2}\xi_2))
\label{hF}
\end{align}
in terms of 
\EQ
\kappa =\frac{\sqrt{c_1}+\sqrt{c_2}}{\sqrt{c_1}-\sqrt{c_2}} >1
\endEQ
where
\EQ
\xi_1 = x-c_1 t-x_1, \quad
\xi_2 = x-c_2 t-x_2
\label{movingcoords'}
\endEQ
are moving coordinates 
centered at initial positions $x=x_1$ and $x=x_2$ respectively. 

The asymptotic form of the $2$-soliton solution $u(t,x)=G/F$ 
as $t\rightarrow\pm\infty$ 
is easily derived by the same moving-coordinate expansions 
considered for the mKdV $2$-soliton solution. 
These expansions yield, after neglecting subdominant exponential terms, 
\EQ
u(t,x) \simeq 
\frac{\sqrt{c_1} \exp(i\phi_1+\sqrt{c_1}(\xi_1 -a_1^\pm))}
{1+\exp(2\sqrt{c_1}(\xi_1 -a_1^\pm))} 
=u_1^\pm ,\quad
t\rightarrow\pm\infty ,\quad
\xi_1=\const
\label{cmkdvfastsoliton}
\endEQ
and
\EQ
u(t,x) \simeq 
\frac{\sqrt{c_2} \exp(i\phi_2+\sqrt{c_2}(\xi_2 -a_2^\pm))}
{1+\exp(2\sqrt{c_2}(\xi_2 -a_2^\pm))} 
=u_2^\pm ,\quad
t\rightarrow\pm\infty ,\quad
\xi_2=\const
\label{cmkdvslowsoliton}
\endEQ
where $a_1^\pm,a_2^\pm$ are shifts in the moving coordinates $\xi_1,\xi_2$
given by \eqref{a1} and \eqref{a2} respectively. 

Thus, 
the $2$-soliton solution \eqref{mkdv2soliton} and \eqref{hG}--\eqref{hF}  
asymptotically has the form of a superposition $u\simeq u_1^\pm+u_2^\pm$ 
for $t\rightarrow \pm\infty$, 
where $u_1^\pm$ is a fast soliton \eqref{cmkdv1soliton}
with speed $c_1$ and phase $\phi_1$
and where $u_2^\pm$ is a slow soliton \eqref{cmkdv1soliton}
with speed $c_2$ and phase $\phi_2$,
whose positions are determined by the shifted moving coordinates 
\eqref{shiftfastcoord} and \eqref{shiftslowcoord}. 
This solution has the same conserved 
momentum \eqref{mkdv2solitonp}, energy \eqref{mkdv2solitone}, 
Galilean energy \eqref{mkdv2solitongale}, 
and center of momentum \eqref{mkdv2solitoncom}
as the mKdV $2$-soliton solution.

\subsection{Asymptotic position shifts}
\noindent\newline\indent
For the $2$-soliton solution $u$ of the Hirota equation, 
the asymptotic expansions \eqref{cmkdvfastsoliton} and \eqref{cmkdvslowsoliton}
for $t\rightarrow-\infty$ compared to $t\rightarrow+\infty$
show that the fast soliton $u_1$ retains its shape, speed and phase, 
but gets shifted forward in position by 
\EQ
\Delta x_1 = a_1^+ - a_1^- 
=\frac{2}{\sqrt{c_1}} \ln\left(\frac{\sqrt{c_1}+\sqrt{c_2}}{\sqrt{c_1}-\sqrt{c_2}}\right)
% 2\ln(\kappa)/\sqrt{c_1} 
>0
\label{hfastshift}
\endEQ
while the slow soliton $u_2$ similarly retains its shape, speed and phase, 
but gets shifted backward in position by 
\EQ
\Delta x_2 = a_2^+ - a_2^- 
= -\frac{2}{\sqrt{c_2}} \ln\left(\frac{\sqrt{c_1}+\sqrt{c_2}}{\sqrt{c_1}-\sqrt{c_2}}\right)
%-2\ln(\kappa)/\sqrt{c_2} 
<0
\label{hslowshift}
\endEQ
These expressions are the same asymptotic shifts 
seen for the collision of mKdV solitons.
In particular, 
the shifts \eqref{hfastshift} and \eqref{hslowshift} here
are independent of the phases of the fast and slow solitons
and also satisfy the center of momentum relations 
\eqref{mkdvshifteq} and \eqref{mkdvu1u2separation}.

\subsection{Interaction profile}
\noindent\newline\indent
In the same way as for the mKdV $2$-soliton solution, 
we will now write the Hirota $2$-soliton solution 
given by \eqref{mkdv2soliton} and \eqref{hG}--\eqref{hF} 
in the equivalent rational-cosh form 
\EQ
u(t,x) = 
\frac{\kappa( \sqrt{c_1}\exp(i\phi_1)\cosh(\theta_2) 
+ \sqrt{c_2}\exp(i\phi_2)\cosh(\theta_1) )}
{(\kappa^2-1)\cos(\phi_1-\phi_2) +\kappa^2 \cosh(\theta_1-\theta_2) 
+ \cosh(\theta_1 +\theta_2)}
\label{h2soliton}
\endEQ
with 
\EQ
\theta_1 = \sqrt{c_1}\xi_1 ,\quad
\theta_2 = \sqrt{c_2}\xi_2 
\endEQ
By shifting the centers of the moving coordinates \eqref{movingcoords'}
to the positions
\EQ
x_1=x_0-c_1 t_0 =0,\quad
x_2=x_0-c_2 t_0 =0, 
\endEQ
we then see that the resulting solution \eqref{h2soliton} of the Hirota equation 
is invariant under the combined space-time reflection 
$x\rightarrow -x$, $t\rightarrow -t$. 
As a consequence, 
this solution $u$ describes a collision of fast and slow solitons 
$u_1^\pm$ and $u_2^\pm$
whose separation will be a minimum at time $t=0$ 
when the amplitude $|u|$ is an even function of $x$. 
This can be understood to be the moment of greatest nonlinear interaction 
between the fast and slow solitons in the collision. 
The shape of $|u|$ at $t=0$ therefore defines 
the interaction profile of the $2$-soliton solution $u$. 
Since this profile is even in $x$, 
its shape is characterized by the convexity of $|u(0,x)|$ at $x=0$. 
An explicit calculation of the convexity yields
\EQ
|u(0,x)|_{xx}\big|_{x=0} =
\frac{(\sqrt{c_1} -\sqrt{c_2})^2
( 4c_1 c_2 -c_1^2-c_2^2 + \sqrt{c_1 c_2}(c_1 +c_2)\cos(\phi_1-\phi_2) )}
{2( c_1+c_2 +2\sqrt{c_1 c_2}\cos(\phi_1-\phi_2) )^{3/2}}
%\frac{1}{2}\frac{(\sqrt{c_1} -\sqrt{c_2})^2
%( \frac{1}{2}\sqrt{c_1 c_2}(\sqrt{\cos^2(\phi_1-\phi_2) +24} -\cos(\phi_1-\phi_2) ) +c_1 +c_2 )
%( \frac{1}{2}\sqrt{c_1 c_2}(\sqrt{\cos^2(\phi_1-\phi_2) +24}+\cos(\phi_1-\phi_2) ) -c_1 -c_2 )}
%{( c_1+c_2 +2\sqrt{c_1 c_2}\cos(\phi_1-\phi_2) )^{3/2}}
\label{h2solitonconvexity}
\endEQ

The sign of the convexity \eqref{h2solitonconvexity} 
depends only on the ratio of speeds and the relative phase 
\EQ
r=c_1/c_2 ,\quad
\Delta\phi=\phi_1-\phi_2 \mod 2\pi
\endEQ
of the fast and slow solitons. 
In particular, since $c_1>c_2>0$, we have
\EQ
\sigma= \sgn(|u(0,x)|_{xx}\big|_{x=0}) 
=\sgn( 4r-r^2 -1 +\sqrt{r}(r+1)\cos\Delta\phi )
\label{hconvexitysgn}
\endEQ
with $r>1$ and $-\pi\leq\Delta\phi\leq \pi$. 
This expression \eqref{hconvexitysgn} is a quartic polynomial in $\sqrt{r}$.
By factorizing 
\begin{align*}
&4r-r^2 -1 +\sqrt{r}(r+1)\cos\Delta\phi = \\&\qquad
\big( \tfrac{1}{2}(\cos\Delta\phi+\sqrt{\cos^2\Delta\phi +24})\sqrt{r}-r-1 \big)
\big( \tfrac{1}{2}(\cos\Delta\phi+\sqrt{\cos^2\Delta\phi +24})\sqrt{r}+r+1 \big)
\end{align*}
and using the inequality 
\begin{align*}
\sqrt{\cos^2\Delta\phi +24} > \pm \cos\Delta\phi
\end{align*}
we obtain 
\EQ
\sigma= \sgn( \tfrac{1}{2}(\cos\Delta\phi+\sqrt{\cos^2\Delta\phi +24})\sqrt{r}-r-1 )
\label{hconvexitysgn'}
\endEQ
which is a quadratic polynomial in $\sqrt{r}$ with two real positive roots
\EQ
\sqrt{r_\pm} = 
\tfrac{1}{4}(\cos\Delta\phi+\sqrt{\cos^2\Delta\phi +24})
\pm\sqrt{ (\tfrac{1}{4}(\cos\Delta\phi+\sqrt{\cos^2\Delta\phi +24}))^2 -1 } 
\ >0
\endEQ
These roots satisfy $r_+ r_- =1$, 
where the case $r_+=r_- =1$ occurs iff $|\Delta\phi|=\pi$.
Hence we have 
\EQ
r_+\geq 1\geq r_-
\endEQ
in all cases. 
Thus the convexity sign \eqref{hconvexitysgn'} is determined by 
\EQ
\sigma= \sgn(\sqrt{r_+}-\sqrt{r})
\label{h2solitonconvexitysgn}
\endEQ
which is indefinite such that 
\begin{align}
\sigma 
\begin{cases}
<0 & \mathtext{ if } r> r_+ \\
>0 & \mathtext{ if } r< r_+ \\
=0 & \mathtext{ if } r= r_+ \\
\end{cases}
\label{hcrit}
\end{align}
where
\EQ
r_+ =
2+\tfrac{1}{4}\big(\cos\Delta\phi+\sqrt{\cos^2\Delta\phi +24}\ \big)
\big( \cos\Delta\phi+\sqrt{\tfrac{1}{2}(\cos\Delta\phi+\sqrt{\cos^2\Delta\phi +24})\cos\Delta\phi +2}\ \big)
\label{sscritspeed}
\endEQ

The result \eqref{hcrit} implies that the $2$-soliton interaction profile 
$|u(0,x)|$ will have either a single peak at $x=0$ if $c_1/c_2 > r_+$
or a double peak around $x=0$ if $c_1/c_2 < r_+$,
as determined by the critical speed ratio \eqref{sscritspeed}
in terms of the relative phase angle $\Delta\phi$. 

\begin{figure}[!h]
\centering\includegraphics[scale=0.85]{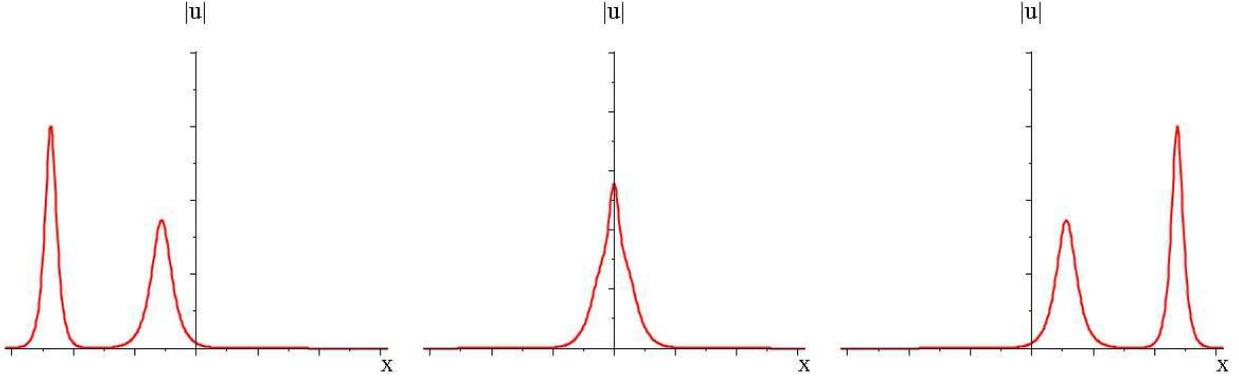}
\caption{Hirota equation $2$-soliton interaction with 
$c_1/c_2=3$, $\phi_1-\phi_2=0.855\pi$}
\label{hirota-merge1}
\end{figure}

\begin{figure}[!h]
\centering\includegraphics[scale=0.85]{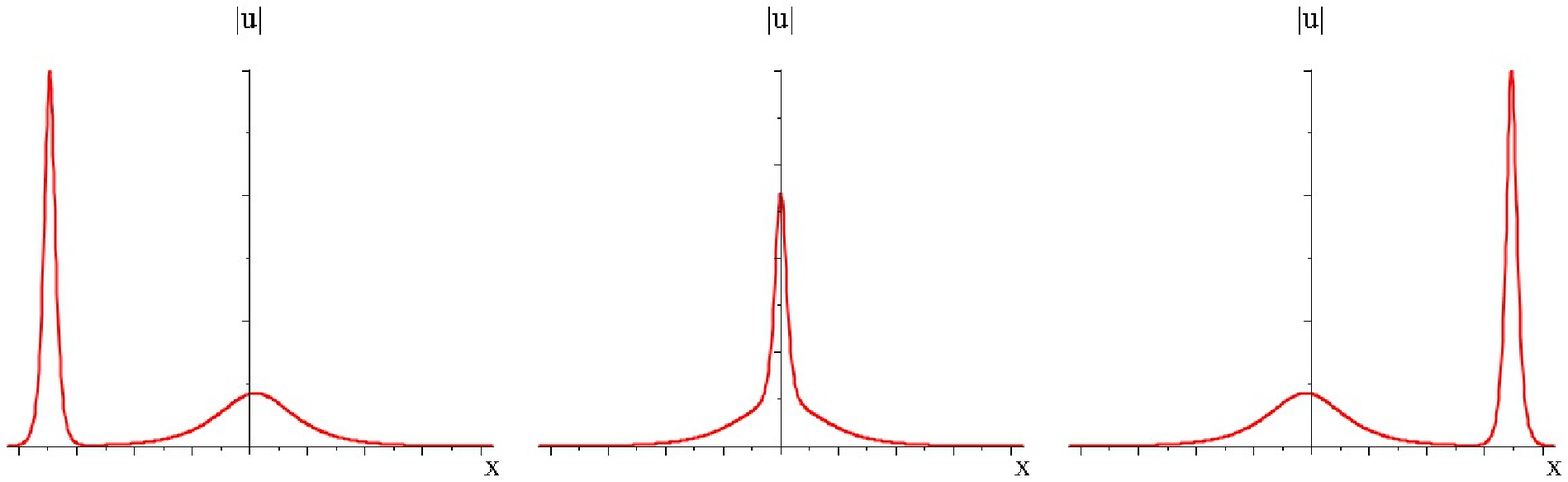}
\caption{Hirota equation $2$-soliton interaction with 
$c_1/c_2=50$, $\phi_1-\phi_2=0.3\pi$}
\label{hirota-merge2}
\end{figure}

\begin{figure}[!h]
\centering\includegraphics[scale=0.85]{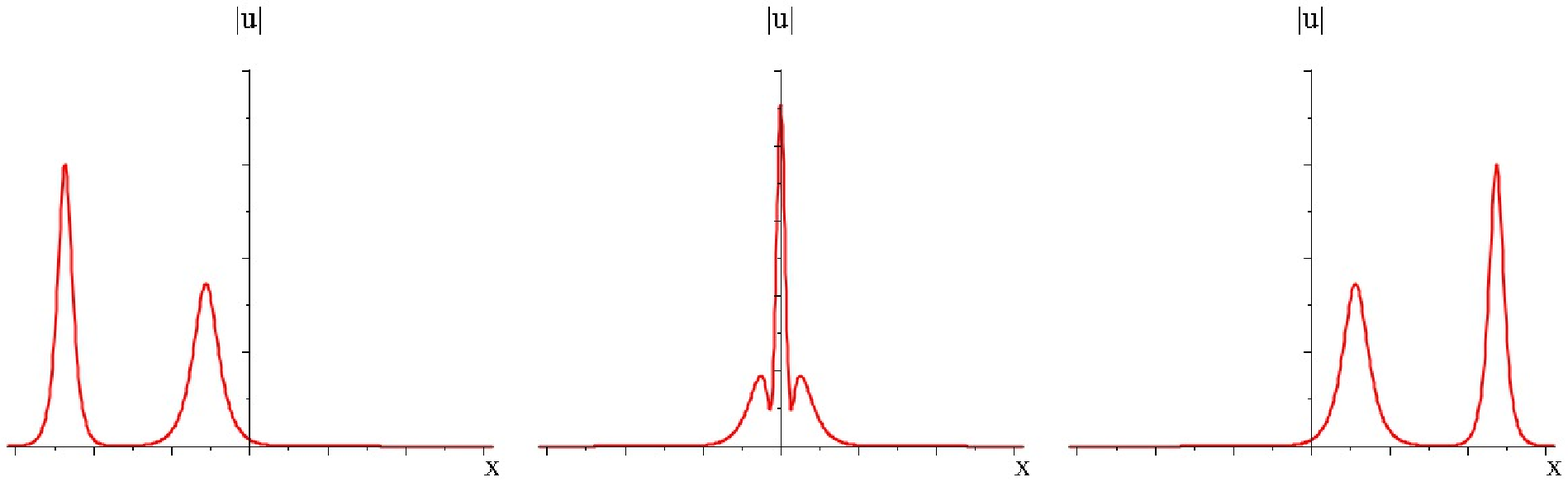}
\caption{Hirota equation $2$-soliton interaction with 
$c_1/c_2=3$, $\phi_1-\phi_2=0.855\pi$}
\label{hirota-absorb1}
\end{figure}

\begin{figure}[!h]
\centering\includegraphics[scale=0.85]{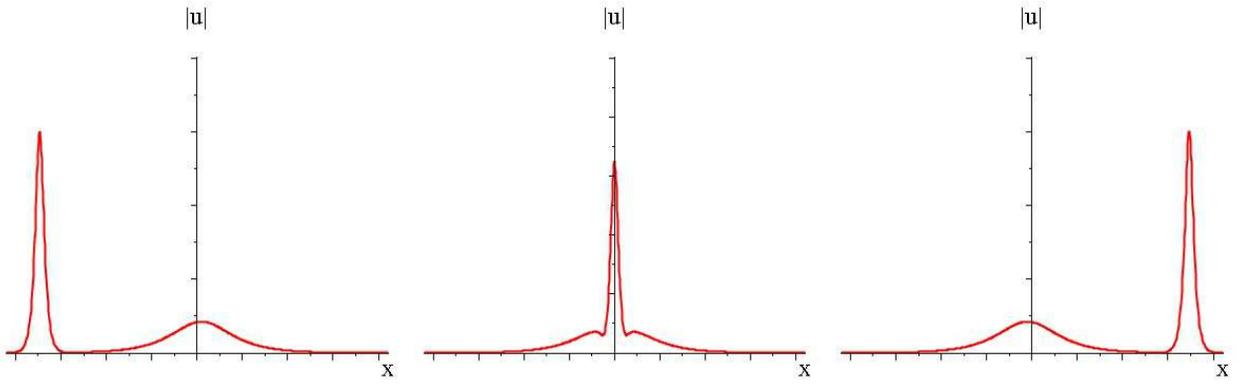}
\caption{Hirota equation $2$-soliton interaction with 
$c_1/c_2=50$, $\phi_1-\phi_2=0.855\pi$}
\label{hirota-absorb2}
\end{figure}

In the case of a single peak,
the profile has one of two different shapes 
depending on whether $|\Delta\phi|$ is greater than or less than 
a certain critical value given by 
some function of $c_1/c_2$ that is determined by the conditions 
$|u(0,x)|_{xx} = |u(0,x)|_{x}=0$
for existence of a saddle point at some $x\neq 0$
(which we can solve for numerically). 
For $|\Delta\phi|$ below the critical value,  
the shape of $|u(0,x)|$ is simply a single peak 
with an exponentially diminishing tail. 
In this case the fast and slow solitons undergo a merge-split interaction, \ie/
where they first merge together at $t=x=0$ and then split apart:
See \figref{hirota-merge1} and \figref{hirota-merge2}.
For $|\Delta\phi|$ above the critical value, 
the shape of $|u(0,x)|$ consists of a pair of side peaks 
around the main peak at $x=0$.
In this case the fast and slow solitons undergo an absorb-emit interaction,
\ie/ where the slow soliton gradually is first absorbed 
by the front side of the fast soliton
and is then emitted from the back side of the fast soliton. 
See \figref{hirota-absorb1} and \figref{hirota-absorb2}.


\begin{figure}[!h]
\centering\includegraphics[scale=0.85]{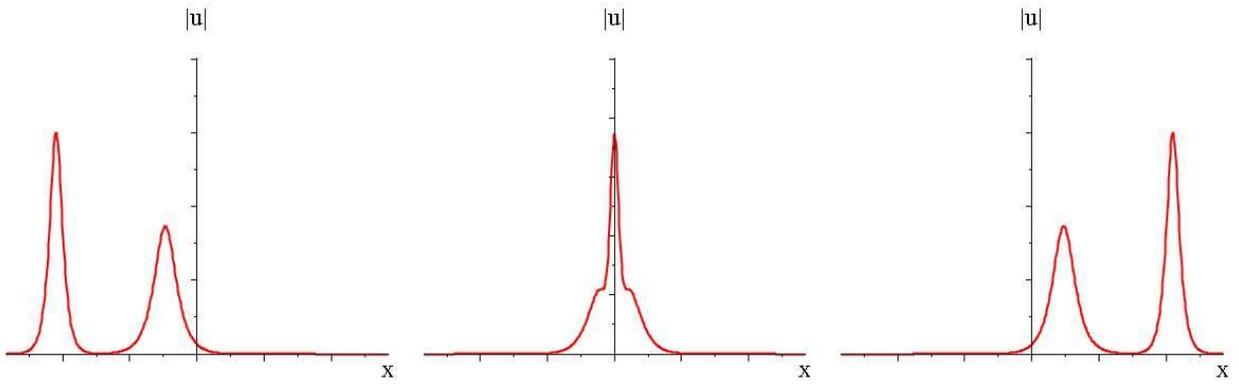}
\caption{Hirota equation $2$-soliton interaction with 
$c_1/c_2=3$, $\phi_1-\phi_2=0.85598\pi$}
\label{h-crit}
\end{figure}

The interaction in the special case
when $|\Delta\phi|$ equals the critical value is shown in \figref{h-crit}. 

\begin{figure}[!h]
\centering\includegraphics[scale=0.85]{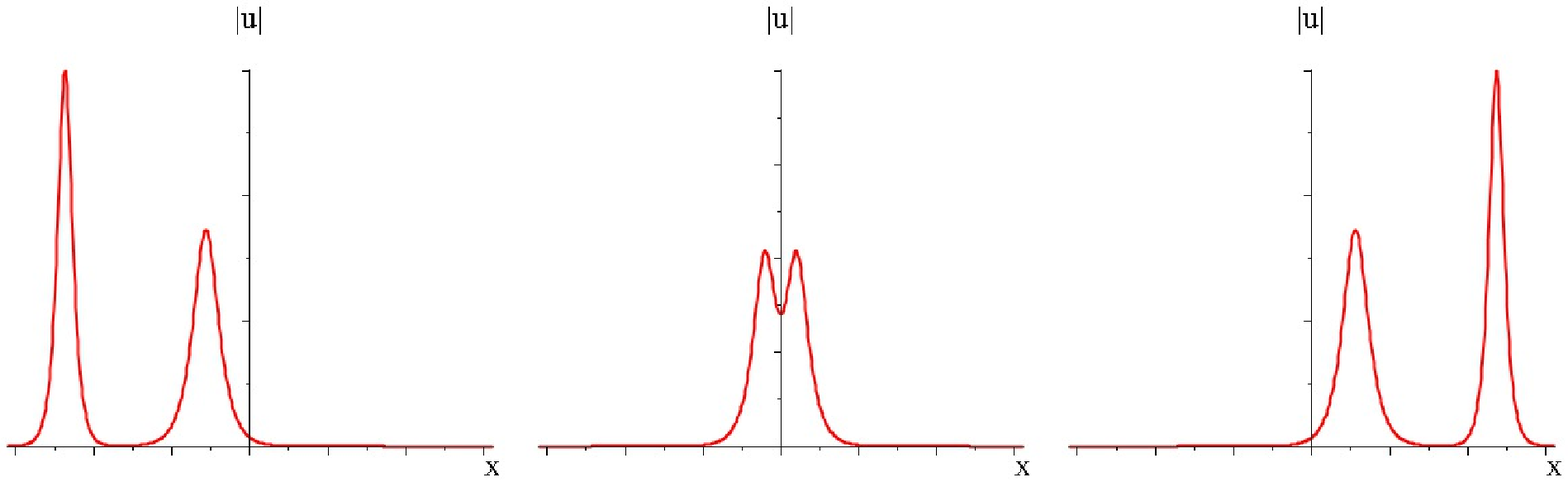}
\caption{Hirota equation $2$-soliton interaction with 
$c_1/c_2=3$, $\phi_1-\phi_2=0.3\pi$}
\label{hirota-bounce}
\end{figure}

In contrast, in the case of a double peak, 
the profile $|u(0,x)|$ always has an exponentially diminishing tail,
regardless of the relative phase angle $|\Delta\phi|$. 
This case describes the fast and slow solitons undergoing 
a bounce-exchange interaction, \ie/ 
where they first bounce and then exchange shapes and speeds at $t=x=0$:
See \figref{hirota-bounce}. 

\begin{figure}[!h]
\centering\includegraphics[scale=0.72]{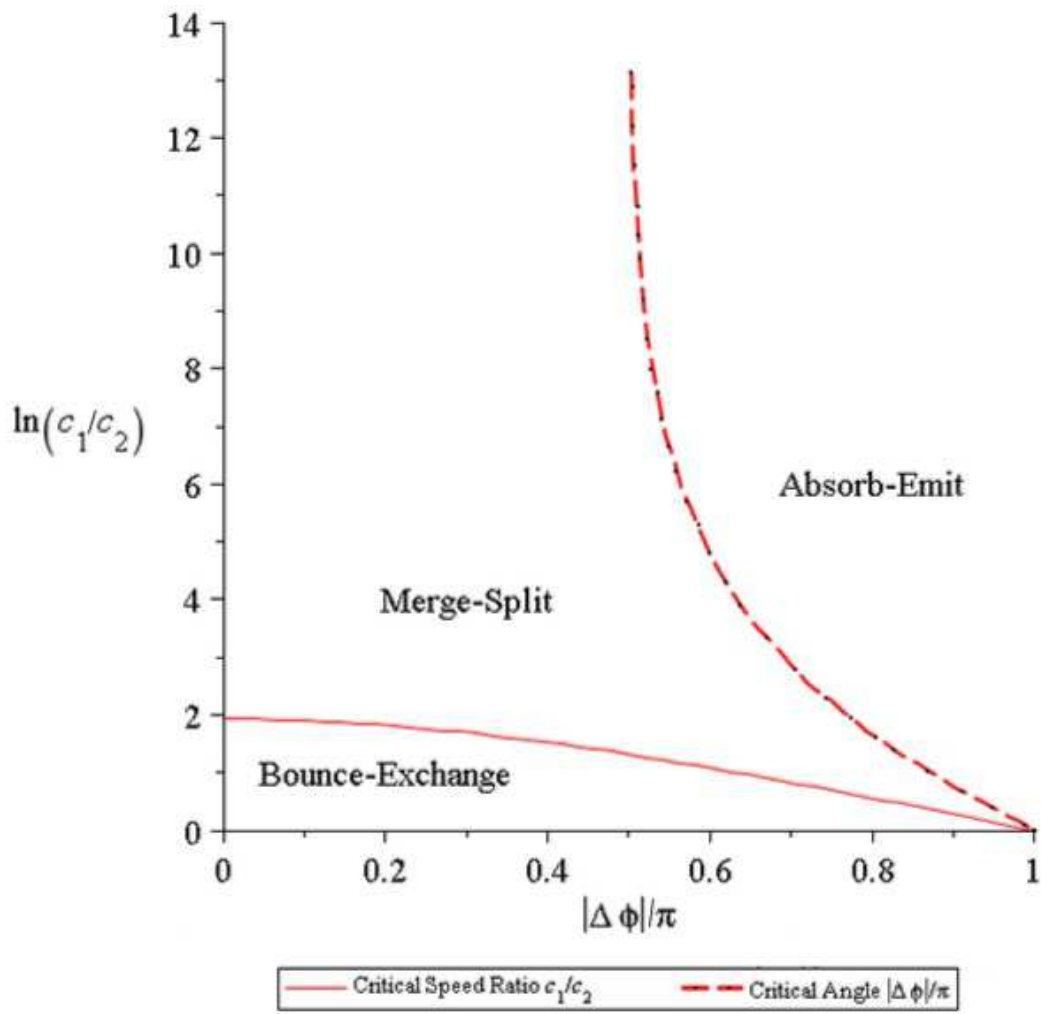}
\caption{Types of $2$-soliton interactions for the Hirota equation}
\label{hcritvalues}
\end{figure}

The range of values of $c_1/c_2$ and $|\Delta\phi|$ that characterize
these three different types of interaction is shown in \figref{hcritvalues}. 

\clearpage

\section{Sasa-Satsuma-mKdV soliton collisions}
\label{ssmkdvcollision} 
For the Sasa-Satsuma equation
\EQ
u_t +6( u\bar u_x +3u_x\bar u )u + u_{xxx} =0
\label{ss}
\endEQ
we begin by remarking \cite{conslaws} that 
it has the same conserved integrals as the Hirota equation
for momentum \eqref{cp}, energy \eqref{ce}, and Galilean energy \eqref{cGale}, 
which define constants of motion for 
all smooth solutions $u(t,x)$ with asymptotic decay $u\rightarrow 0$
as $x\rightarrow\pm\infty$. 
In addition these integrals have the same relation to 
the center of momentum \eqref{ccom}--\eqref{ccomrel}
that holds for the Hirota equation. 

The $1$-soliton solution for the Sasa-Satsuma equation has the same form 
\eqref{cmkdv1soliton}--\eqref{movingcoord'} 
as the Hirota $1$-soliton with speed $c>0$ and phase $-\pi\leq \phi\leq \pi$.
We recall that this solution describes 
a stable uni-directional travelling wave $u$
whose amplitude $|u|$ is the same as the amplitude of the mKdV solitary wave \eqref{mkdv1soliton}, 
and hence $u$ also has the same constants of motion \eqref{cp}--\eqref{cGale} 
and center of momentum \eqref{ccom} as those of the mKdV solitary wave. 
Thus, 
the position of the peak amplitude of $u$
coincides with the center of momentum $\mathcal{X}(t)= ct$
which can be shifted arbitrarily via 
a space translation $x\rightarrow x-x_0$ 
applied to the moving coordinate \eqref{movingcoord'}, 
yielding
\EQ
u(t,x) = \frac{\sqrt{c}}{2}\exp(i\phi) \sech\left(\sqrt{c}\xi\right)
= \frac{\sqrt{c} \exp(i\phi+\sqrt{c}\xi)}{1+\exp(2\sqrt{c}\xi)} 
\label{ss1soliton}
\endEQ
with 
\EQ
\xi = x-ct-x_0
\label{ssmovingcoord}
\endEQ

\subsection{$2$-soliton solution}
\noindent\newline\indent
The $2$-soliton solution of the Sasa-Satsuma equation 
describing collisions where a fast soliton with speed $c_1$ and phase $\phi_1$ 
overtakes a slow soliton with speed $c_2$ and phase $\phi_2$
has not appeared previously in an explicit form
\cite{GilHieNimOht}. 
We give a simple derivation of this solution in the appendix, 
based on using a rational-exponential ansatz similar to the form of 
the $2$-soliton solution of the Hirota equation. 
This derivation yields
\EQ
u(t,x) = \frac{G}{F}
\label{ss2soliton}
\endEQ
with 
\begin{align}
G =& \frac{\rho}{\sqrt{\gamma}} \Big(
\sqrt{c_1}\exp(i\phi_1+\sqrt{c_1}\xi_1)
\big( (\sqrt{c_1}+\sqrt{c_2}\exp(2i(\phi_2-\phi_1)))\exp(2\sqrt{c_2}\xi_2) +\gamma \big)
\nonumber\\&\qquad
+ \sqrt{c_2}\exp(i\phi_2+ \sqrt{c_2}\xi_2)
\big( \gamma \exp(2\sqrt{c_1}\xi_1)+ \sqrt{c_2}+\sqrt{c_1}\exp(2i(\phi_1-\phi_2)) \big)
\Big) 
\label{ssG}\\
F=& 1+ 2 \cos(\phi_1-\phi_2) ((\sqrt{c_1}+\sqrt{c_2})\rho^2-1) \exp(\sqrt{c_1}\xi_1 +\sqrt{c_2}\xi_2) 
\nonumber\\&\qquad
+\gamma\rho^2( \exp(2\sqrt{c_1}\xi_1) + \exp(2\sqrt{c_2}\xi_2) )
+ \exp(2(\sqrt{c_1}\xi_1 +\sqrt{c_2}\xi_2))
\label{ssF}
\end{align}
in terms of 
\EQ
\rho =\frac{\sqrt{\sqrt{c_1}+\sqrt{c_2}}}{\sqrt{c_1}-\sqrt{c_2}} >0
\endEQ
and
\EQ
\gamma=|\sqrt{c_1}\exp(i(\phi_2-\phi_1)) + \sqrt{c_2}\exp(i(\phi_1-\phi_2))|
= \sqrt{c_1 + c_2 +2\sqrt{c_1 c_2}\cos 2(\phi_1-\phi_2)} >0
\endEQ
where
\EQ
\xi_1 = x-c_1 t-x_1, \quad
\xi_2 = x-c_2 t-x_2
\label{movingcoords''}
\endEQ
are moving coordinates 
centered at initial positions $x=x_1$ and $x=x_2$ respectively. 
(See the animations of collisions at 
http://lie.math.brocku.ca/\url{~}sanco/solitons/sasa-satsuma.php
)

We will now examine the asymptotic form of the $2$-soliton solution 
\eqref{ss2soliton}--\eqref{ssF} as $t\rightarrow\pm\infty$ 
by means of the same moving-coordinate expansions 
used for the mKdV $2$-soliton solution. 

First we hold $\xi_1$ fixed and asymptotically expand $F$ and $G$ 
for large $\zeta=t\Delta c-\Delta x$ with $\xi_2=\xi_1+\zeta$. 
This yields, after neglecting subdominant exponential terms, 
\begin{align}
& G\simeq
\begin{cases}
\dfrac{\rho}{\sqrt{\gamma}} \sqrt{c_1} \exp(i\phi_1 +(\sqrt{c_1} +2\sqrt{c_2})\xi_1) ( \sqrt{c_1} + \sqrt{c_2}\exp(2i(\phi_2 -\phi_1)) )
\exp(2\sqrt{c_2}\zeta) 
& \zeta\rightarrow +\infty\\
\rho\sqrt{\gamma} \sqrt{c_1} \exp(i\phi_1 +\sqrt{c_1}\xi_1)
& \zeta\rightarrow -\infty
\end{cases}
\\
& F\simeq
\begin{cases}
\big( \gamma\rho^2 \exp(2\sqrt{c_2}\xi_1) + \exp(2(\sqrt{c_1} +\sqrt{c_2})\xi_1) \big)
\exp(2\sqrt{c_2}\zeta) 
& \zeta\rightarrow +\infty\\
1+\gamma\rho^2 \exp(2\sqrt{c_1}\xi_1) 
& \zeta\rightarrow -\infty
\end{cases}
\end{align}
and hence we obtain the expansion
\begin{align}
& u\simeq 
\begin{cases}
\dfrac{(\sqrt{c_1}/\rho\sqrt{\gamma^3}) \exp(i\phi_1 +\sqrt{c_1}\xi_1)
( \sqrt{c_1} + \sqrt{c_2}\exp(2i(\phi_2 -\phi_1)) )}
{1+(1/\rho^2\gamma)\exp(2\sqrt{c_1}\xi_1)} 
& \zeta\rightarrow +\infty\\
\dfrac{\sqrt{c_1} \rho\sqrt{\gamma} \exp(i\phi_1 +\sqrt{c_1}\xi_1)}
{1+\rho^2\gamma \exp(2\sqrt{c_1}\xi_1)} 
& \zeta\rightarrow -\infty
\end{cases}
\label{ssu1}
\end{align}
where $\zeta\rightarrow\pm\infty$ corresponds to $t\rightarrow\pm\infty$. 
In the asymptotic past, 
this expansion \eqref{ssu1} has the form of a $1$-soliton solution 
\EQ
u \simeq 
\frac{\sqrt{c_1} \exp(i\phi_1+\sqrt{c_1}(\xi_1 -a_1^-))}
{1+\exp(2\sqrt{c_1}(\xi_1 -a_1^-))} 
=u_1^- ,\quad
t\rightarrow -\infty ,\quad
\xi_1=\const
\label{ssfastsoliton-}
\endEQ
in which the moving coordinate $\xi_1$ is shifted by 
\EQ
a_1^- = -\ln(\rho\sqrt{\gamma})/\sqrt{c_1}
\label{ssa1-}
\endEQ
Similarly, 
in the asymptotic future, 
the expansion \eqref{ssu1} again has the form of a $1$-soliton solution 
\EQ
u \simeq 
\frac{\sqrt{c_1} \exp(i(\phi_1+\nu_1)+\sqrt{c_1}(\xi_1 -a_1^+))}
{1+\exp(2\sqrt{c_1}(\xi_1 -a_1^+))} 
=u_1^+ ,\quad
t\rightarrow +\infty ,\quad
\xi_1=\const
\label{ssfastsoliton+}
\endEQ
in which the moving coordinate $\xi_1$ is now shifted by 
\EQ
a_1^+ = \ln(\rho\sqrt{\gamma})/\sqrt{c_1}
\label{ssa1+}
\endEQ
while in addition there is a phase shift given by 
\EQ
\frac{\sqrt{c_1} + \sqrt{c_2}\exp(2i(\phi_2 -\phi_1))}{\gamma}
=\exp(i\nu_1)
\endEQ
where
\EQ
\nu_1 =\arctan\left(
\dfrac{\sqrt{c_2}\sin 2(\phi_1-\phi_2)}{\sqrt{c_1}+ \sqrt{c_2}\cos 2(\phi_1-\phi_2)}\right)
\label{ssfastphaseshift}
\endEQ

Second we hold $\xi_2$ fixed and asymptotically expand $F$ and $G$ 
for large $\zeta=t\Delta c-\Delta x$ with $\xi_1=\xi_2-\zeta$. 
After neglecting subdominant exponential terms, we obtain 
\begin{align}
& G\simeq
\begin{cases}
\dfrac{\rho}{\sqrt{\gamma}} \sqrt{c_2} \exp(i\phi_2 +\sqrt{c_2}\xi_2)
\big( \sqrt{c_2} + \sqrt{c_1}\exp(2i(\phi_1 -\phi_2)) \big)
& \zeta\rightarrow +\infty\\
\rho\sqrt{\gamma} \sqrt{c_2} \exp(i\phi_2 +(\sqrt{c_2} +2\sqrt{c_1})\xi_2) 
\exp(-2\sqrt{c_1}\zeta) 
& \zeta\rightarrow -\infty
\end{cases}
\\
& F\simeq
\begin{cases}
1+\gamma\rho^2 \exp(2\sqrt{c_2}\xi_2) 
& \zeta\rightarrow +\infty\\
\big( \gamma\rho^2 \exp(2\sqrt{c_1}\xi_2) + \exp(2(\sqrt{c_1} +\sqrt{c_2})\xi_2) \big)
\exp(-2\sqrt{c_1}\zeta) 
& \zeta\rightarrow -\infty
\end{cases}
\end{align}
which yields the expansion
\begin{align}
& u\simeq 
\begin{cases}
\dfrac{(\sqrt{c_2}/\rho/\sqrt{\gamma}) \exp(i\phi_2 +\sqrt{c_2}\xi_2)
( \sqrt{c_2} + \sqrt{c_1}\exp(2i(\phi_1 -\phi_2)) )}
{1+\rho^2\gamma\exp(2\sqrt{c_2}\xi_2)} 
& \zeta\rightarrow +\infty\\
\dfrac{(\sqrt{c_2}/ \rho\sqrt{\gamma}) \exp(i\phi_2 +\sqrt{c_2}\xi_2)}
{1+(1/\rho^2\gamma) \exp(2\sqrt{c_2}\xi_2)} 
& \zeta\rightarrow -\infty
\end{cases}
\label{ssu2}
\end{align}
where $\zeta\rightarrow\pm\infty$ corresponds to $t\rightarrow\pm\infty$. 
In the asymptotic past, 
the expansion \eqref{ssu2} has the form of a $1$-soliton solution 
\EQ
u \simeq 
\frac{\sqrt{c_2} \exp(i\phi_2+\sqrt{c_2}(\xi_2 -a_2^-))}
{1+\exp(2\sqrt{c_2}(\xi_2 -a_2^-))} 
=u_2^- ,\quad
t\rightarrow -\infty ,\quad
\xi_2=\const
\label{ssslowsoliton-}
\endEQ
in which the moving coordinate $\xi_2$ is shifted by 
\EQ
a_2^- = \ln(\rho\sqrt{\gamma})/\sqrt{c_2}
\label{ssa2-}
\endEQ
In the asymptotic future, 
this expansion \eqref{ssu2} similarly has the form of a $1$-soliton solution 
\EQ
u \simeq 
\frac{\sqrt{c_2} \exp(i(\phi_2+\nu_2)+\sqrt{c_2}(\xi_2 -a_2^+))}
{1+\exp(2\sqrt{c_2}(\xi_2 -a_2^+))} 
=u_2^+ ,\quad
t\rightarrow +\infty ,\quad
\xi_2=\const
\endEQ
in which the moving coordinate $\xi_2$ is now shifted by 
\EQ
a_2^+ = -\ln(\rho\sqrt{\gamma})/\sqrt{c_2}
\label{ssa2+}
\endEQ
while in addition there is a phase shift given by 
\EQ
\frac{\sqrt{c_2} + \sqrt{c_1}\exp(2i(\phi_1 -\phi_2))}{\gamma}
=\exp(i\nu_2)
\endEQ
where
\EQ
\nu_2 =\arctan\left(
\dfrac{\sqrt{c_1}\sin 2(\phi_2-\phi_1)}{\sqrt{c_2}+ \sqrt{c_1}\cos 2(\phi_2-\phi_1)}\right)
\label{ssslowphaseshift}
\endEQ

Thus, 
for $t\rightarrow \pm\infty$, 
the $2$-soliton solution \eqref{ss2soliton}--\eqref{ssF} 
asymptotically has the form of a superposition $u\simeq u_1^\pm+u_2^\pm$ 
of a fast soliton $u_1^\pm$ and a slow soliton $u_2^\pm$,
with speeds $c_1$ and $c_2$. 
As a consequence, this solution has the same conserved 
momentum \eqref{mkdv2solitonp}, energy \eqref{mkdv2solitone}, 
Galilean energy \eqref{mkdv2solitongale}, 
and center of momentum \eqref{mkdv2solitoncom}
as the mKdV $2$-soliton solution.

\subsection{Asymptotic position and phase shifts}
\noindent\newline\indent
In the $2$-soliton solution $u$ of the Sasa-Satsuma equation, 
the positions of the fast soliton $u_1^\pm$ and the slow soliton $u_2^\pm$ 
in the asymptotic past ($t\rightarrow-\infty$) and future ($t\rightarrow+\infty$)
are determined by the shifted moving coordinates 
$\xi_1^\pm = \xi_1-a_1^\pm$ and $\xi_2^\pm = \xi_2-a_2^\pm$. 
We thus see that the fast soliton $u_1$ retains its shape and speed, 
but gets shifted forward in position by 
\EQ
\Delta x_1 = a_1^+ - a_1^- 
=\frac{1}{\sqrt{c_1}} \ln\left(
\frac{(\sqrt{c_1}+\sqrt{c_2})\sqrt{c_1 + c_2 +2\sqrt{c_1 c_2}\cos 2(\phi_1-\phi_2)}}
{(\sqrt{c_1}-\sqrt{c_2})^2}
\right)
>0
\label{ssfastshift}
\endEQ
while the slow soliton $u_2$ similarly retains its shape and speed, 
but gets shifted backward in position by 
\EQ
\Delta x_2 = a_2^+ - a_2^- 
= -\frac{1}{\sqrt{c_2}} \ln\left(
\frac{(\sqrt{c_1}+\sqrt{c_2})\sqrt{c_1 + c_2 +2\sqrt{c_1 c_2}\cos 2(\phi_1-\phi_2)}}
{(\sqrt{c_1}-\sqrt{c_2})^2}
\right)
<0
\label{ssslowshift}
\endEQ
These asymptotic shifts satisfy the relation 
\EQ
\sqrt{c_1}\Delta x_1 + \sqrt{c_2}\Delta x_2 =0
\endEQ
which can be understood as a consequence of the motion of 
the center of momentum of the $2$-soliton solution $u$
in the same way as for the Hirota equation. 
Interestingly, in contrast to collisions of Hirota solitons, 
here the shifts \eqref{ssfastshift} and \eqref{ssslowshift} 
depend on the relative phase angle $\phi_1-\phi_2$ 
between the fast and slow solitons in the collision. 

Even more interestingly, in the collision, 
both the fast and slow solitons undergo a shift in phase 
given by \eqref{ssfastphaseshift} and \eqref{ssslowphaseshift} respectively.
The features of these asymptotic shifts can be understood from
the reflection properties of the $2$-soliton solution as follows. 
First we write this solution in the equivalent form 
\EQ
u(t,x) = 
\frac{\rho\sqrt{\gamma}  \big(
\sqrt{c_1}\exp(i\phi_1)( \exp(-\theta_2) + \exp(i\nu_1+\theta_2) )
+ \sqrt{c_2}\exp(i\phi_2)( \exp(\theta_1) + \exp(i\nu_2-\theta_1) ) \big)}
{2\big( 
\cosh(\theta_1 +\theta_2) +\rho^2 \gamma \cosh(\theta_1-\theta_2) 
+ \cos(\phi_1-\phi_2) ((\sqrt{c_1}+\sqrt{c_2})\rho^2-1) \big)}
\label{ss2soliton'}
\endEQ
with 
\EQ
\theta_1 =  \sqrt{c_1}(x-c_1 t) ,\quad
\theta_2 =  \sqrt{c_2}(x-c_2 t) ,
\endEQ
where we have used a space-time translation 
$x\rightarrow x-x_0$ and $t\rightarrow t-t_0$ 
to shift the centers of the moving coordinates \eqref{movingcoords''}
to the positions 
\EQ
x_1=x_0-c_1 t_0 =0,\quad
x_2=x_0-c_2 t_0 =0
\endEQ
The solution \eqref{ss2soliton'} of the Sasa-Satsuma equation 
then exhibits an invariance
\EQ
u(-t,-x)=\exp(i\nu){\bar u}(t,x)
\label{ssinvariance}
\endEQ
where the phase factor is given by 
\EQ
\exp(i\nu)= \frac{\sqrt{c_2}\exp(i\phi_1) + \sqrt{c_1}\exp(i\phi_2)}{\gamma}
= \exp(i(\nu_1 +2\phi_1)) = \exp(i(\nu_2 +2\phi_2))
\endEQ
in terms of the asymptotic phase shifts $\nu_1$ and $\nu_2$. 
Next we shift the phases 
\EQ
\phi_1 \rightarrow \phi_1 -\nu/2 ,\quad
\phi_2 \rightarrow \phi_2 -\nu/2 
\endEQ
which corresponds to an overall phase rotation of the solution
\EQ
u\rightarrow \exp(-i\nu/2) u =\hat u
\endEQ
This does not affect the relative phase angle $\phi_1-\phi_2$ 
between the fast and slow solitons in the collision,
whereby the resulting solution 
\EQ
{\hat u}(t,x) = 
\frac{\rho \sqrt{\gamma} \big( 
\sqrt{c_1}\cosh(\theta_2+i\nu_1/2)+ \sqrt{c_2}\cosh(\theta_1-i\nu_2/2) 
\big)}
{\big( 
\cosh(\theta_1 +\theta_2) +\rho^2 \gamma\cosh(\theta_1-\theta_2) 
+ \cos(\phi_1-\phi_2) ((\sqrt{c_1}+\sqrt{c_2})\rho^2-1) 
\big)}
\label{ssinv2soliton}
\endEQ
is invariant under space-time reflection $x\rightarrow -x$, $t\rightarrow -t$,
combined with phase conjugation, 
\EQ
{\hat u}(-t,-x)=\bar{\hat u}(t,x)
\label{ssreflection}
\endEQ

For $t\rightarrow\pm\infty$, 
this $2$-soliton solution $\hat u$ of the Sasa-Satsuma equation 
describes a collision where, 
in the asymptotic past ($t\rightarrow -\infty$), 
a fast soliton $\hat u_1^-$ with speed $c_1$, 
phase $\phi_1^-=\phi_1 -\nu/2=-nu_1/2$,
and center of momentum $\chi_1^-(t)=a_1^- +c_1 t$ 
overtakes a slow soliton $\hat u_2^-$ with speed $c_2$, 
phase $\phi_2^-=\phi_2 -\nu/2=-\nu_2/2$,
and center of momentum $\chi_2^-(t)=a_2^- +c_2 t$. 
In the asymptotic future ($t\rightarrow +\infty$),
the fast soliton $\hat u_1^+$ undergoes a shift in both 
position $\chi_1^+(t)=a_1^+ +c_1 t$ 
and phase $\phi_1^+=\nu_1 + \phi_1^- =\nu_1/2$,
while the slow soliton $\hat u_2^+$ similarly undergoes both 
a position shift $\chi_2^+(t)=a_2^+ +c_2 t$ 
and a phase shift $\phi_2^+=\nu_2 + \phi_2^- =\nu_2/2$,
where these phase shifts are related by
\EQ 
\phi_1^- -\phi_2^- = -(\phi_1^+ -\phi_2^+)
\label{ssphaseshift}
\endEQ
due to the reflection property \eqref{ssreflection}. 

This asymptotic phase relation \eqref{ssphaseshift} is equivalent to 
\EQ
\nu_1-\nu_2 = -2(\phi_1-\phi_2)
\endEQ
Thus, surprisingly, 
the relative phase angle between the fast and slow solitons 
is not preserved in the collision but instead changes sign.

\subsection{Interaction profile}
\noindent\newline\indent
The invariance property \eqref{ssinvariance} of the $2$-soliton solution $u$
shows that the separation in positions of the peak amplitude of 
the fast and slow solitons $u_1^\pm$ and $u_2^\pm$ in the collision 
will be a minimum at time $t=0$ 
when the amplitude $|u|$ is an even function of $x$. 
The shape of $|u|$ at $t=0$ thereby defines 
the interaction profile of the $2$-soliton solution $u$,
which can be understood to be the moment of greatest nonlinear interaction 
between the fast and slow solitons in the collision. 
This profile is characterized by the convexity of $|u(0,x)|$ at $x=0$. 

By an explicit calculation, we find that the convexity is given by 
\begin{align}
&|u(0,x)|_{xx}\big|_{x=0} =
\nonumber\\&\qquad
\frac{ (\sqrt{c_1} -\sqrt{c_2})^2 \sqrt{2c_1 c_2} 
( C c_1 c_2 + B\sqrt{c_1 c_2} (c_1 +  c_2) -A(c_1^2 +c_2^2) )}
{\sqrt{\gamma( K(\sqrt{c_1} +\sqrt{c_2}+\gamma) -L(\sqrt{c_1} +\sqrt{c_2}) )}
( Q(\sqrt{c_1} +\sqrt{c_2}) -P(\sqrt{c_1} +\sqrt{c_2}-\gamma) ) }
\label{ss2solitonconvexity}
\end{align}
where
\begin{align}
& A=\cos^2\Delta\phi +3 \cos\Delta\phi +3 \\
& B=\cos^2\Delta\phi +6 \cos\Delta\phi \\
& C=18 \cos\Delta\phi +10 
\end{align}
and
\begin{align}
& K= c_1+c_2 +2 \sqrt{c_1 c_2} \cos(\phi_1-\phi_2) \\
& L= 2 \sqrt{c_1 c_2}( 1-\cos^2(\phi_1-\phi_2) ) \\
& P= c_1+c_2 +4\sqrt{c_1 c_2} \\
& Q= 2 \sqrt{c_1 c_2}( 4+3\cos(\phi_1-\phi_2) ) 
\end{align}
The sign of the convexity \eqref{ss2solitonconvexity} 
depends only on the ratio of speeds and the relative phase 
\EQ
r=c_1/c_2, \quad
\Delta\phi=\phi_1-\phi_2 \mod 2\pi
\endEQ
of the fast and slow solitons. 
Since we have $c_1>c_2>0$, $\pi\geq\phi_1\geq-\pi$, $\pi\geq\phi_2\geq-\pi$, 
these parameters are restricted to the respective intervals
\EQ
r>1, \quad
\pi \geq|\Delta\phi|\geq 0
\label{range}
\endEQ
To determine the conditions under which the convexity is positive or negative,
we will separately consider the signs of the factors 
in the numerator and denominator expressions. 

The sign of the numerator in the convexity \eqref{ss2solitonconvexity} 
is given by 
\EQ
\sigma= \sgn( Cr + B\sqrt{r}(r+1) -A(r^2 +1) )
\label{ssconvexitynumer}
\endEQ
This expression can be factorized
\EQ
Cr + B\sqrt{r}(r+1) -A(r^2 +1) 
= A\big( ((D+B)/2A)\sqrt{r} -r -1 \big)\big( ((D-B)/2A)\sqrt{r} +r +1 \big)
\label{ssfactornumer}
\endEQ
by means of the identity 
\EQ
(D-B)(D+B) = 8A(\cos^2\Delta\phi +12\cos\Delta\phi +8)
\label{ssident}
\endEQ
where
\EQ
D=(3\cos\Delta\phi +4)\sqrt{\cos^2\Delta\phi +12 \cos\Delta\phi +12}
\endEQ
We note this identity \eqref{ssident} also implies the inequality 
\EQ
D > B
\label{ssinequal}
\endEQ
as follows. 
By evaluating $D-B$ for $\Delta\phi=\arccos(-6+2\sqrt{7})$, 
which is the sole real root of the right-hand side of the identity, 
we find $D-B=8(3\sqrt{7}-7)$ is positive. 
This implies $\sgn(D-B) >0$ holds for all values of $\Delta\phi$ 
in the interval \eqref{range}, 
since $D-B$ may change sign only at a value 
where the identity \eqref{ssident} vanishes. 

To continue, from the inequality \eqref{ssinequal}, 
we see 
\EQ
((D-B)/2A)\sqrt{r} +r +1 >0
\endEQ
holds throughout the interval \eqref{range}, 
and hence the factorization \eqref{ssconvexitynumer}--\eqref{ssfactornumer}
yields
\EQ
\sigma = \sgn( ((D-B)/2A)\sqrt{r} +r +1 )
\label{ssnumersgn}
\endEQ
which determines the sign of the numerator. 

The sign of the denominator in the convexity \eqref{ss2solitonconvexity} 
is determined by 
\EQ
\delta = \sgn( 2E\sqrt{r}(\sqrt{r}+1) -(r+1 +4\sqrt{r})(\sqrt{r}+1-\gamma) )
\endEQ
where
\EQ
E=3\cos\Delta\phi +4
\endEQ
This sign can be evaluated in terms of 
the roots of the right-hand side of the identity 
\begin{align}
&\big( 2E(r+\sqrt{r}) -(r+1 +4\sqrt{r})(\sqrt{r}+1-\gamma) \big)
\big( 2E(r+\sqrt{r}) -(r+1 +4\sqrt{r})(\sqrt{r}+1+\gamma) \big)
\nonumber\\\quad
&= \sqrt{r} \big( 4r(\sqrt{r}-1)^4(1-\cos\Delta\phi)^4 \big)^2 (A(r^2 +1)-B\sqrt{r}(r+1)- Cr)
\end{align}
There are two real roots contained in the interval \eqref{range},
which are the only values of $\Delta\phi$ and $r$ 
where $\delta$ may change sign. 
For the root $\Delta\phi=0$, 
we find $\delta=\sgn( 14\sqrt{r}(\sqrt{r}+1) )$ is positive since $r>1$. 
Similarly, for the other root given by $r+1+((D-B)/2A)\sqrt{r}=0$,
we find $\delta$ is again positive. 
Hence this implies 
\EQ
\delta >0
\label{ssdenomsgn}
\endEQ
holds throughout the interval \eqref{range}. 
We remark that the same argument can be used to show that 
the expression inside the square-root factor 
in the denominator of the convexity \eqref{ss2solitonconvexity} 
is positive for all values of $\Delta\phi$ and $r$ in this interval. 

Therefore, 
the signs of the numerator \eqref{ssnumersgn} and the denominator \eqref{ssdenomsgn}
yield the sign of the convexity 
\EQ
\sgn(|u(0,x)|_{xx}\big|_{x=0}) =\sigma
\label{ssconvexitysgn}
\endEQ
From expression \eqref{ssnumersgn} 
this sign is a quadratic polynomial in $\sqrt{r}$ with two roots
\EQ
\sqrt{r_\pm} = ((B+D)/4A) \pm\sqrt{ ((B+D)/4A)^2 -1 } 
\endEQ
which satisfy $r_+ r_- =1$. 
The two roots are real and positive iff 
$((B+D)/4A)^2 \geq 1$. 
This condition can be expressed as $D^2 \geq (4A-B)^2$ 
which simplifies to 
\EQ
(6\cos\Delta\phi+1)(\cos^2\Delta\phi+3\cos\Delta\phi+3)\geq 0
\endEQ
Hence, the condition for $r_+$ and $r_-$ to be real and positive is 
$\cos\Delta\phi\geq -1/6$. 
In this case the roots have the property $r_+\geq 1\geq r_-$ 
such that $r_+=r_- =1$ when $\Delta\phi=\arccos(-1/6)$.

\begin{figure}[!h]
\centering\includegraphics[scale=0.80]{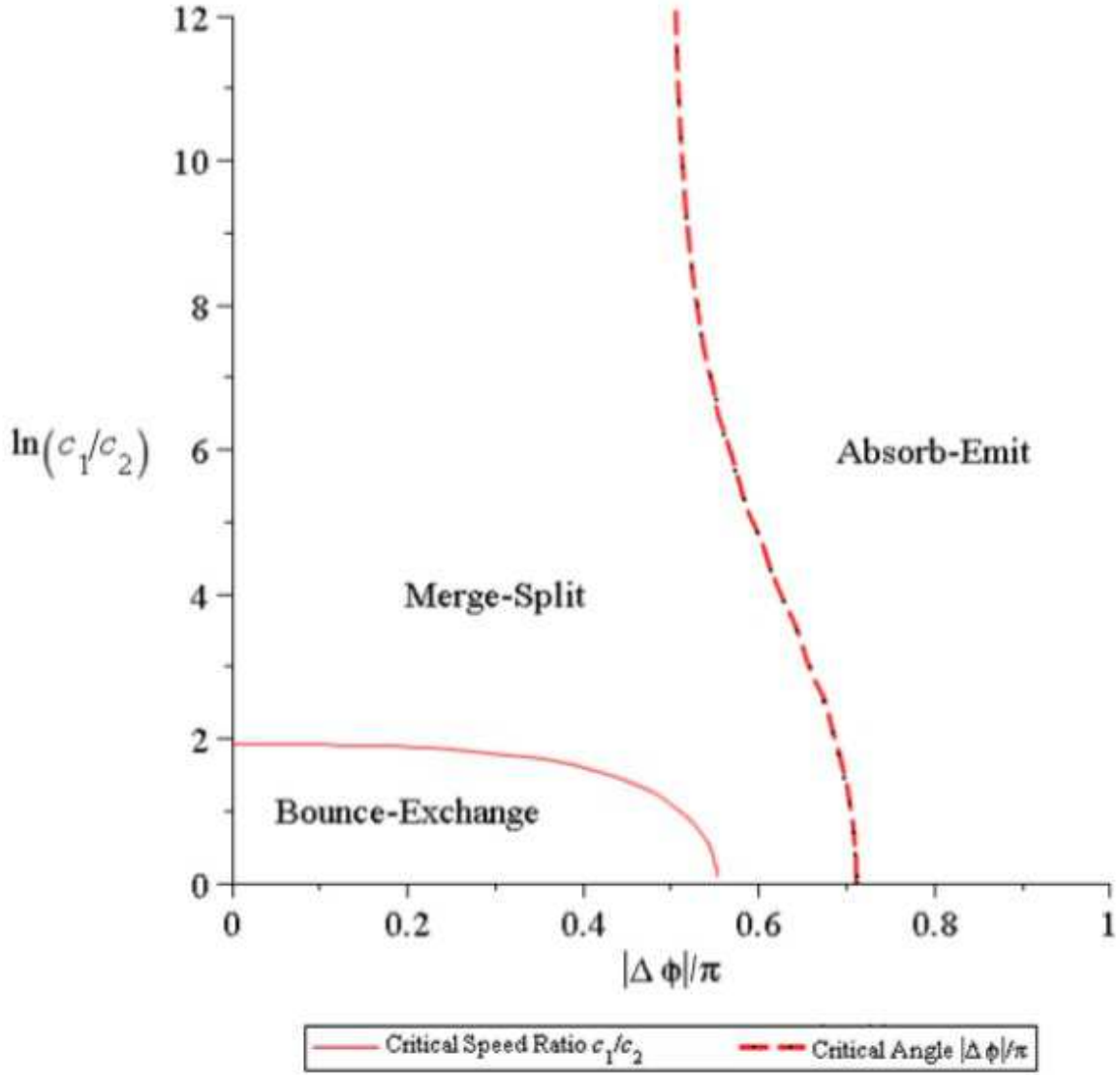}
\caption{Types of $2$-soliton interactions for the Sasa-Satsuma equation}
\label{sscritvalues}
\end{figure}

As a result, 
since $r>1$, 
the convexity sign \eqref{ssconvexitysgn} is given by 
\begin{align}
\sigma 
\begin{cases}
= \sgn(\sqrt{r_+}-\sqrt{r}) &\mathtext{ if } 
|\Delta\phi| \leq \arccos(-1/6) \approx 0.55\pi \\
< 0 &\mathtext{ if } 
|\Delta\phi| \geq \arccos(-1/6) \approx 0.55\pi
\end{cases}
\label{sscrit}
\end{align}
where
\EQ
r_+ =
\frac{C}{2A}+\frac{B+D}{4A^2}\big(B+\sqrt{B(B+D)/2 +A(C-2A)}\big)
\endEQ
Thus when the relative phase angle $|\Delta\phi|$ is greater than 
$\arccos(-1/6) \approx 0.55\pi$, 
the $2$-soliton interaction profile $|u(0,x)|$ will always 
have a single peak at $x=0$,  
whereas when the relative phase angle $|\Delta\phi|$ is less than 
$\arccos(-1/6) \approx 0.55\pi$, 
the profile $|u(0,x)|$ will have either a single peak at $x=0$ 
if the speed ratio is $c_1/c_2 > r_+$ 
or a double peak around $x=0$ if the speed ratio is $c_1/c_2 < r_+$. 
Additionally, for a double peak, 
the profile will always have an exponentially diminishing tail.
For a single peak, 
the profile instead will have either a pair of side peaks around $x=0$
or just an exponentially diminishing tail, 
depending on whether $|\Delta\phi|$ is greater than or less than 
a certain critical value for which the profile exhibits 
a saddle point at some $x\neq 0$
as determined by the conditions $|u(0,x)|_{xx} = |u(0,x)|_{x}=0$
(which we can solve numerically). 
The critical angle $|\Delta\phi|$ 
and the critical speed ratio $c_1/c_2=r_+$
are shown in \figref{sscritvalues}. 

\begin{figure}[!h]
\centering
\includegraphics[scale=0.85]{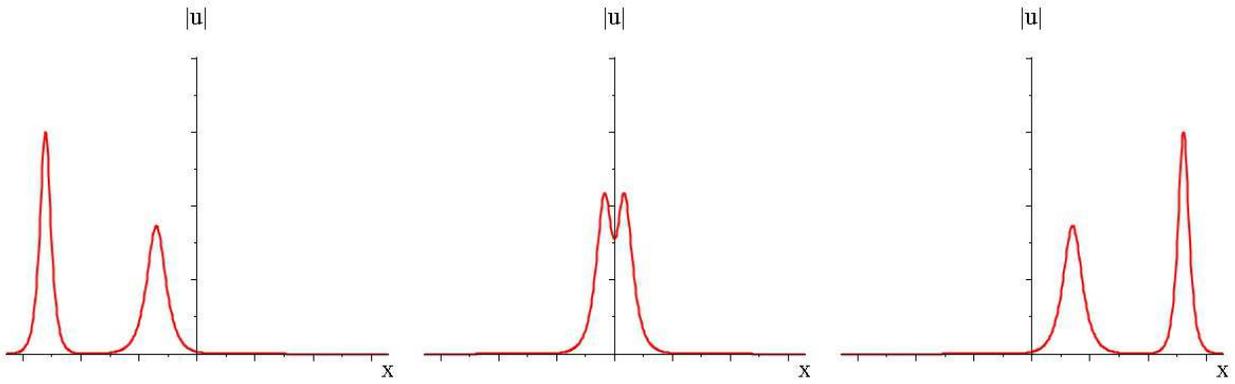}
\caption{Sasa-Satsuma equation $2$-soliton interaction with 
$c_1/c_2=3$, $\phi_1-\phi_2=0.35\pi$}
\label{ss-bounce}
\end{figure}

In the case of a double peak profile, 
the interaction is a bounce-exchange, \ie/
where the fast and slow solitons first bounce 
and then exchange shapes and speeds at $t=x=0$. See \figref{ss-bounce}. 

\begin{figure}[!h]
\centering
\includegraphics[scale=0.85]{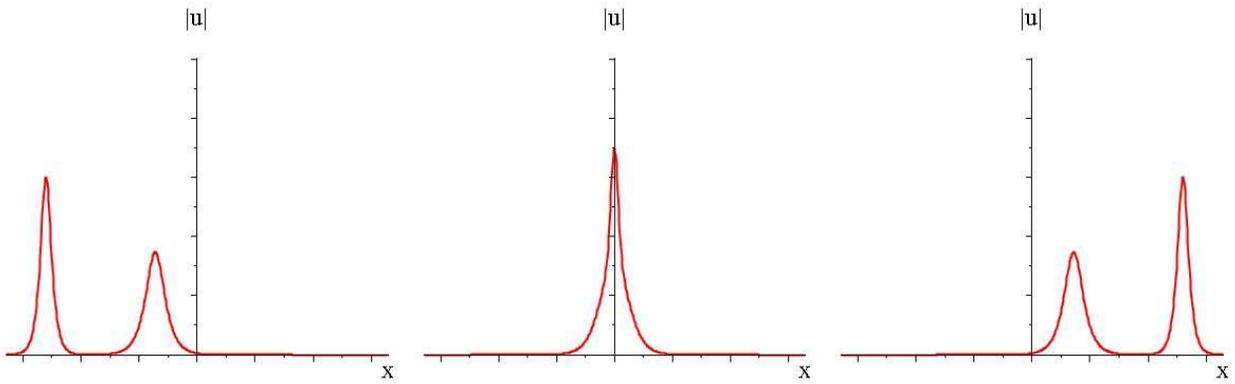}
\caption{Sasa-Satsuma equation $2$-soliton interaction with 
$c_1/c_2=3$, $\phi_1-\phi_2=0.85\pi$}
\label{ss-merge1}
\end{figure}

\begin{figure}[!h]
\centering
\includegraphics[scale=0.85]{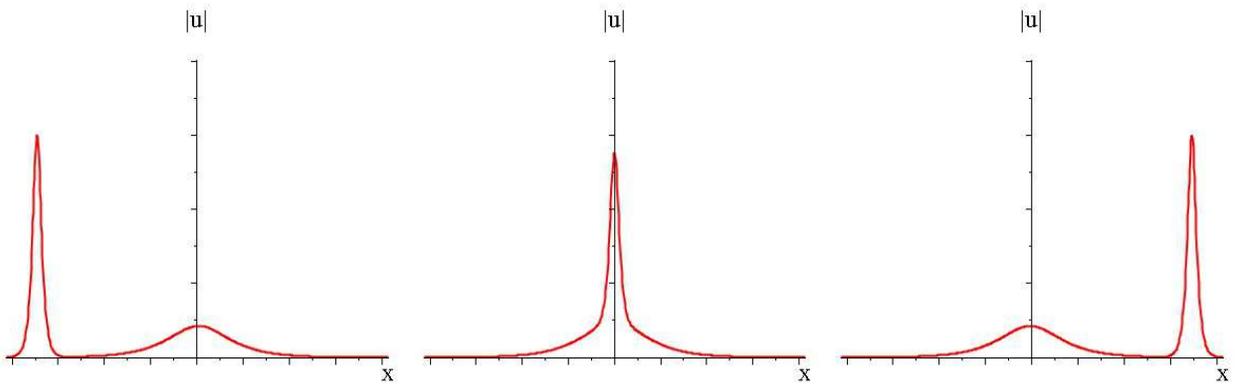}
\caption{Sasa-Satsuma equation $2$-soliton interaction with 
$c_1/c_2=50$, $\phi_1-\phi_2=0.35\pi$}
\label{ss-merge2}
\end{figure}

In the case of a single peak profile without side peaks, 
the interaction of the fast and slow solitons is a merge-split, \ie/
where the solitons first merge together at $t=x=0$ and then split apart. 
See \figref{ss-merge1} and \figref{ss-merge2}. 

\begin{figure}[!h]
\centering
\includegraphics[scale=0.85]{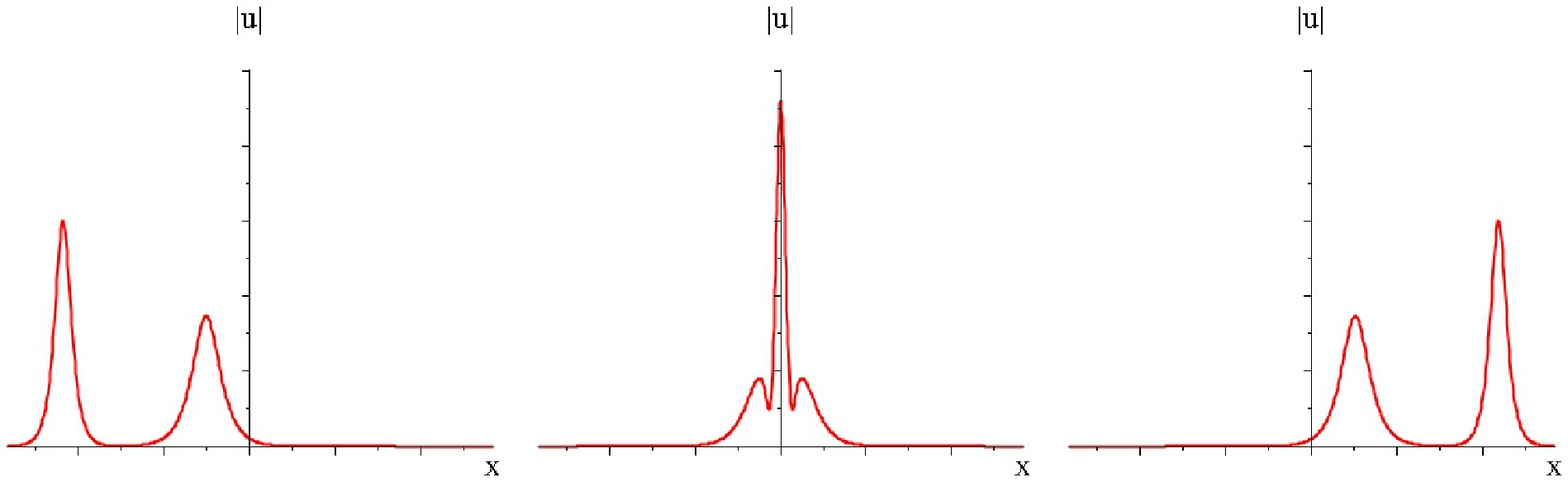}
\caption{Sasa-Satsuma equation $2$-soliton interaction with 
$c_1/c_2=3$, $\phi_1-\phi_2=0.855\pi$}
\label{ss-absorb1}
\end{figure}

\begin{figure}[!h]
\centering
\includegraphics[scale=0.85]{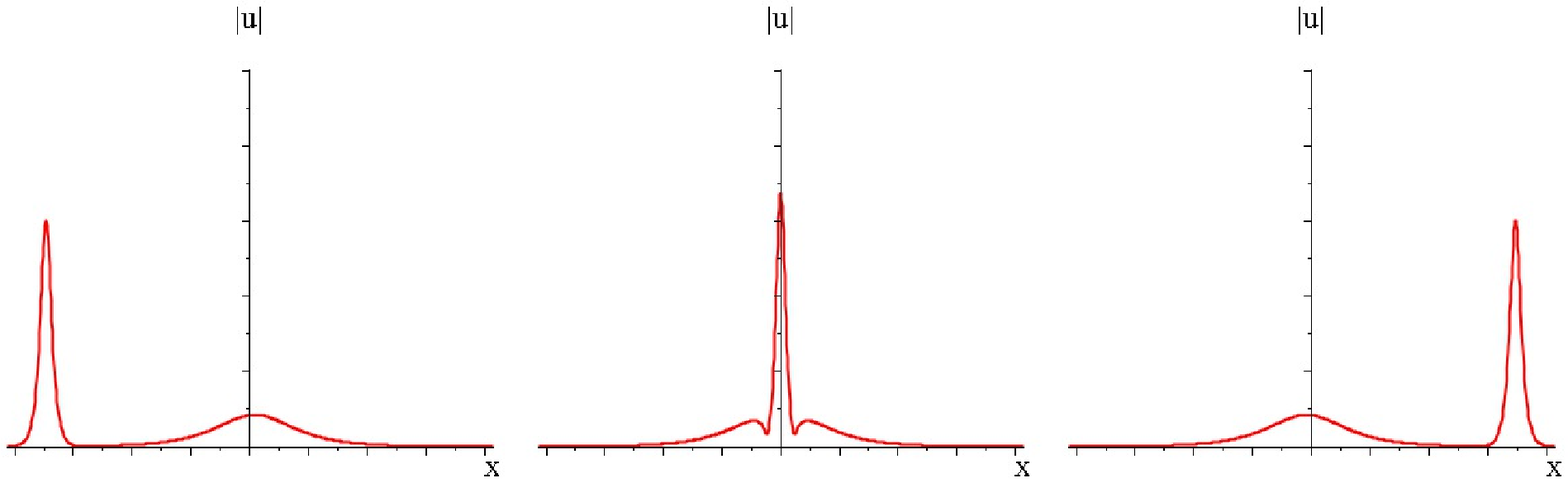}
\caption{Sasa-Satsuma equation $2$-soliton interaction with 
$c_1/c_2=50$, $\phi_1-\phi_2=0.855\pi$}
\label{ss-absorb2}
\end{figure}

The case with side peaks is an absorb-emit interaction,
\ie/ where the slow soliton gradually is first absorbed 
by the front side of the fast soliton
and is then emitted from the back side of the fast soliton. 
See \figref{ss-absorb1} and \figref{ss-absorb2}.

\begin{figure}[!h]
\centering
\includegraphics[scale=0.85]{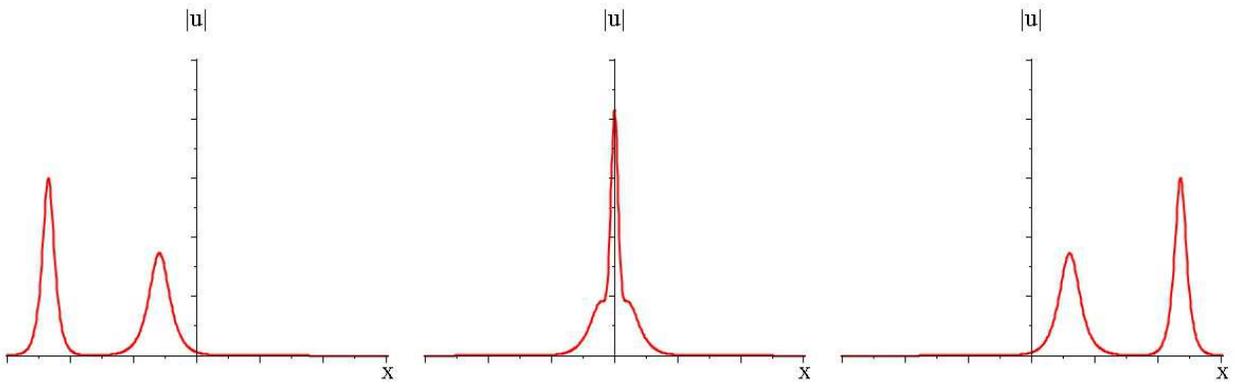}
\caption{Sasa-Satsuma equation $2$-soliton interaction with 
$c_1/c_2=3$, $\phi_1-\phi_2=0.85032\pi$}
\label{ss-crit}
\end{figure}

The special case of a saddle profile is shown in \figref{ss-crit}. 

\clearpage

\section{ Comparison of soliton collisions for the Hirota and Sasa-Satsuma equations }
\label{compare}

Up to a constant phase factor,
the $2$-soliton solution for both the Hirota and Sasa-Satsuma equations
reduces to the mKdV $2$-soliton solution 
when the relative phase angle $\Delta\phi$ between the fast and slow solitons
in the collision is $0$ (same-orientation case) 
or $\pm\pi$ (opposite-orientation case). 

In the Hirota $2$-soliton solution, 
the shift in positions of the fast and slow solitons 
is exactly the same as in the mKdV $2$-soliton solution,
which depends only on the respective speeds $c_1$ and $c_2$ of the solitons. 
Moreover, 
in both the Hirota $2$-soliton solution and the mKdV $2$-soliton solution, 
there is no shift in the phase angles or orientations of the solitons. 
 
\begin{figure}[!h]
\centering
\includegraphics[scale=0.85]{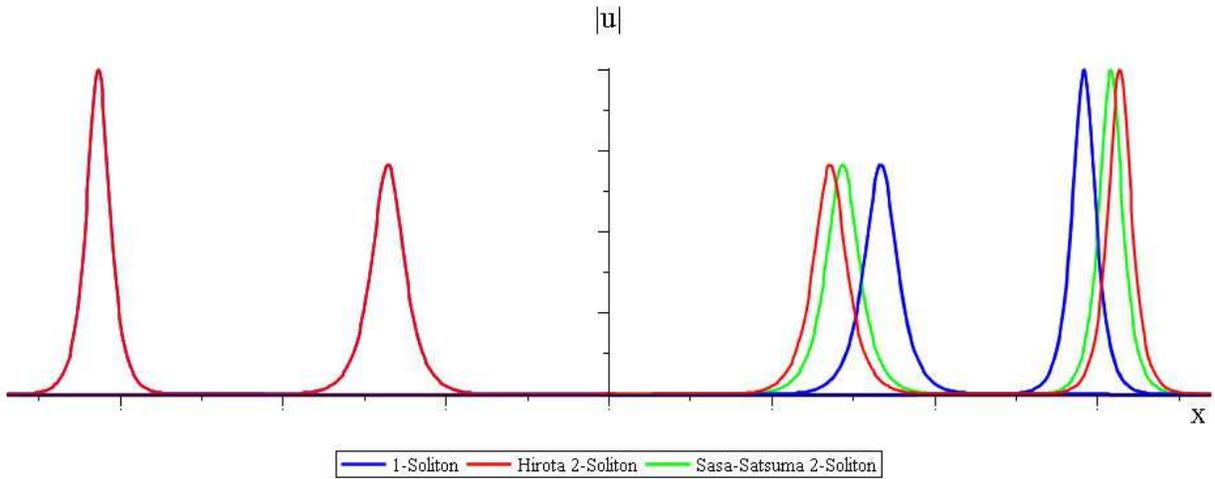}
\caption{Sasa-Satsuma and Hirota $2$-soliton interactions with 
$c_1/c_2=2$, $\phi_1-\phi_2=0.5\pi$}
\label{compare-shifts}
\end{figure}

In contrast, 
the position shifts of the fast and slow solitons 
in the Sasa-Satsuma $2$-soliton solution 
depend on their relative phase angle $\Delta\phi$ in addition to 
their speeds $c_1$ and $c_2$. 
See \figref{compare-shifts}. 
More remarkable is that the collision changes the sign of the relative phase angle.

Both the Sasa-Satsuma and Hirota $2$-soliton solutions describe 
three different types of collisions, 
which are separated by a critical speed ratio and a critical phase angle
shown in \figref{sscritvalues} and \figref{hcritvalues}. 
Interestingly, 
for collisions in the Sasa-Satsuma case, 
the angle $|\Delta\phi|\approx 0.55\pi$ 
at which the critical speed ratio $r_+$ approaches the limit ratio of $1$ 
is strictly less than the critical phase angle $|\Delta\phi|\approx 0.712\pi$
when $c_1/c_2$ approaches the same limit ratio of $1$,
whereas these angles are both equal to $\pi$ 
for collisions in the Hirota case.

\section{ Concluding remarks }
\label{remarks}

Our work in this paper studying 
the interaction properties of complex mKdV solitons
can be extended in at least three interesting directions. 

First, 
the Hirota equation \eqref{h} and the Sasa-Satsuma equation \eqref{ss}
are gauge-equivalent to a third order NLS equation \cite{GilHieNimOht}
\EQ
q_{\tilde t} \pm i \sqrt{\frac{v}{2}}(3 q_{\tilde x\tilde x} + |q|^2 q)
+ \alpha |q|^2 q_{\tilde x} + \beta |q| |q|_{\tilde x} q
+  q_{\tilde x\tilde x\tilde x} =0
\label{nls}
\endEQ
under the Galilean-phase transformation
\EQ
u(t,x) = q(\tilde t,\tilde x)\exp\left(\pm i\sqrt{\frac{v}{2}}(\tilde x - v\tilde t)\right)
\label{nlstransformation}
\endEQ
where $v>0$ is a speed parameter, 
with $\alpha=24$, $\beta=0$ in the Hirota case
and $\alpha=\beta=12$ in the Sasa-Satsuma case. 
Our main results (cf. \secref{compare}) concerning the properties of 
collisions of solitary waves for the Hirota and Sasa-Satsuma equations 
will thus directly carry over to solitary waves of the form
\EQ
q(\tilde t,\tilde x)= 
\frac{\sqrt{c}}{2} \exp\left(i\phi + i\omega\tilde t\right)
\sech\left(\sqrt{c}(\tilde x- \tilde c \tilde t)\right) 
\endEQ
with 
\EQ
\omega =\pm\sqrt{v/2}(c + v/2),\qquad 
\tilde c = c+3v/2 >0
\endEQ
for the corresponding cases of the NLS equation \eqref{nls}.
In particular, 
such solitary waves will exhibit three distinct types of collisions, 
which are separated by a critical speed ratio and a critical phase angle
in terms of the speeds $\tilde c_1$, $\tilde c_2$, 
the phases $\phi_1$, $\phi_2$, and the frequencies $\omega_1$, $\omega_2$
of the solitary waves in the collision. 
For all collisions, 
the waves undergo position shifts in both the Hirota and Sasa-Satsuma cases,
as well as phase shifts in the Sasa-Satsuma case,
such that the shifts depend only on the parameters 
$c_1$, $c_2$, and $\phi_1-\phi_2$. 

Second, as indicated by the correspondence \eqref{nlstransformation}, 
the Hirota and Sasa-Satsuma equations each admit more general 
solitary wave solutions given by the form 
\EQ
u(t,x)= \exp(i\psi(t))f(x-ct)
\endEQ
where 
\EQ
\psi(t)=\omega t+\phi
\endEQ
is a time-varying phase, with constant parameters $\omega,\phi$,
and where $c>0$ is the wave speed. 
Collisions of such waves should display interesting interaction properties
with new features beyond those with $\omega=0$ 
studied in \secref{hmkdvcollision} and \secref{ssmkdvcollision}. 

Second, 
both the Hirota and Sasa-Satsuma equations have a natural multi-component
generalization given by the two known types of 
$O(N)$-invariant integrable mKdV equations \cite{SokWol}
for a $N$-component vector variable $\vec u$. 
In particular, 
under the identification between a $2$-component vector $\vec u=(u_1,u_2)$
and a complex scalar $u=u_1 +i u_2$, 
the vector version of the Hirota equation \eqref{h} is given by 
\EQ
\vec u_t+24 |\vec u|^2\vec u_x +\vec u_{xxx}=0
\label{vecmkdv1}
\endEQ
and the vector version of the Sasa-Satsuma equation \eqref{ss} is given by 
\EQ
\vec u_t+12 |\vec u|^2\vec u_x +12(\vec u\cdot\vec u_x)\vec u +\vec u_{xxx}=0
\label{vecmkdv2}
\endEQ
For all $N\geq 2$, these two vector equations are integrable
and admit vector solitary wave solutions of the form 
\EQ
\vec u(t,x) = f(x-ct)\hat\psi 
\endEQ
with wave speed $c>0$, 
where $\hat\psi$ is an arbitrary constant unit vector
and $f$ is the {\em sech} solitary wave profile 
for the real scalar mKdV equation \eqref{mkdv}. 
In forthcoming work, we plan to generalize the results 
in \secref{hmkdvcollision} and \secref{ssmkdvcollision} 
to study the interaction properties of collisions of vector solitons of 
both equations \eqref{vecmkdv1} and \eqref{vecmkdv2}
for $N>2$,
where the collision involves 
a fast soliton with vector orientation $\hat\psi_1$ and speed $c_1$ 
overtaking a slow soliton with vector orientation $\hat\psi_2$ and speed $c_2$.

\section*{Acknowledgement}
S. Anco is supported by an NSERC research grant. 
The authors thank Takayuki Tsuchida for many valuable comments.

\appendix
\section{}

The $2$-soliton solution of the Sasa-Satsuma equation \eqref{ss}
can be derived most easily by a computational-ansatz version of 
the Hirota method \cite{Hirbook} as follows. 
We first use the standard rational transformation 
\EQ
u=G/F
\label{ssuGF}
\endEQ
with $F$ real and $G$ complex.
This transformation converts the Sasa-Satsuma equation 
into the equivalent rational form 
\EQ
0=
\frac{1}{F^2} \big( D_t(G,F) + D_x^3(G,F) \big)
+ \frac{1}{F^3}  \big( 6G D_x(\bar G,G) \big) 
+ \frac{1}{F^4} \big( 24 |G|^2 -3 D_x^2(F,F)) D_x(G,F) \big)
\label{ssGFeq}
\endEQ
as written in terms of Hirota's bilinear $D$ operator given by 
\begin{align}
& D(a,b)=b Da-a Db \\
& D^2(a,b)=bD^2 a + aD^2 b-2(Da)Db \\
& D^3(a,b)=bD^3 a -3(Db)D^2 a +3(Da)D^2 b -a D^3 b 
\end{align}
Through the introduction of an auxiliary variable $H$, 
there is a natural splitting of equation \eqref{ssGFeq} into 
a bilinear system of equations
\begin{align}
D_x^2(F,F)= 8G\bar G ,\quad
D_t(G,F) + D_x^3(G,F) =iHG ,\quad
iFH= 6 D_x(G,\bar G)
\label{ssGFHeqs}
\end{align}
for the real variables $F,H$ and the complex variable $G$. 

We now observe that the $1$-soliton solution of the bilinear system \eqref{ssGFHeqs}
is given by the simple exponential polynomials
\EQ
G= A\exp(\theta) ,\quad
F= 1+B\exp(2\theta) ,\quad
H=0
\endEQ
with 
\EQ
\theta=kx-\omega t, \quad
\omega=k^3,
\endEQ
where $A$ and $B$ are respectively complex and real parameters 
satisfying the algebraic relation
\EQ
|A|^2 = k^2 B
\endEQ
In particular, 
if we write $A=|A|\exp(i\phi)$ by a polar decomposition
and express $B=(|A|/k)^2$, 
this yields
\EQ
G= k\exp(k\xi) ,\quad
F= 1+\exp(2k\xi) 
\label{ss1solitonGF}
\endEQ
in terms of the moving coordinate
\EQ
\xi=x-k^2 t -x_0
\endEQ
which is centered at the initial position
\EQ
x_0= \ln(k/|A|)/k
\endEQ
Then the rational exponential solution 
obtained from expressions \eqref{ssuGF} and \eqref{ss1solitonGF} 
matches the solitary wave solution \eqref{ss1soliton}--\eqref{ssmovingcoord}
having speed $c=k^2 >0$ and center of momentum $\chi(t)=k^2 t +x_0$. 

To obtain the $2$-soliton solution of the bilinear system \eqref{ssGFHeqs},
we use the ansatz 
\begin{align}
G=& 
A_1\exp(\theta_1) + A_2\exp(\theta_2) 
+ C_1\exp(\theta_1 +2\theta_2) + C_2\exp(\theta_2 +2\theta_1) 
\label{ss2solitonG}\\
F=&
1+B_1\exp(2\theta_1) +B_2\exp(2\theta_2) + I\exp(\theta_1 +\theta_2) 
+ J\exp(2\theta_1 +2\theta_2) 
\label{ss2solitonF}
\end{align}
and 
\EQ
H = K \exp(\theta_1 +\theta_2) 
\label{ss2solitonH}
\endEQ
with 
\EQ
\theta_1=k_1 x-\omega_1 t, \quad
\theta_2=k_2 x-\omega_2 t,
\endEQ
where $B_1,B_2,I,J,K$ are real parameters, 
and $A_1,A_2,C_1,C_2$ are complex parameters. 
In this ansatz, 
the form \eqref{ss2solitonG}--\eqref{ss2solitonF} for the variables $G,F$ 
is motivated by the rational exponential form of 
the mKdV $2$-soliton \eqref{mkdv2soliton},
while the form \eqref{ss2solitonH} for the auxiliary variable $H$ 
comes from balancing the highest-power terms 
in the third equation in the bilinear system \eqref{ssGFHeqs}.
Substitution of $G,F,H$ into the system \eqref{ssGFHeqs} 
leads to three equations, 
which are each a polynomial in the exponentials
$\exp(\theta_1)$ and $\exp(\theta_2)$. 
The coefficients of these polynomials directly yield an overdetermined 
bilinear system of algebraic equations that can be solved for the parameters 
$B_1,B_2,I,J,K,C_1,C_2$ in terms of the pair of complex parameters $A_1,A_2$. 
This gives us the result (obtained by a Maple calculation)
\begin{align}
&
\omega_1 = k_1^3 ,\quad
\omega_2 = k_2^3 ,\quad
\theta_1 = k_1(x-k_1^2 t) ,\quad
\theta_2 = k_2(x-k_2^2 t) ,
\label{sstheta}\\
&
B_1 = \frac{|A_1|^2}{k_1^2} ,\quad
B_2 = \frac{|A_2|^2}{k_2^2} ,
\label{ssB1B2}\\
&
C_1 = 
\frac{(k_1-k_2)(k_1 A_1\bar A_2 -k_2 A_2\bar A_1) A_2}{k_2^2 (k_1 +k_2)^2} ,\quad
C_2 = 
\frac{(k_1-k_2)(k_1 A_1\bar A_2 -k_2 A_2\bar A_1) A_1}{k_1^2 (k_1 +k_2)^2} ,
\label{ssC1C2}\\
&
I = 4\frac{A_1\bar A_2 + A_2\bar A_1}{(k_1 +k_2)^2} ,\quad
J= \frac{(k_1-k_2)^2(k_1 A_2\bar A_1 -k_2 A_1\bar A_2) (k_1 A_1\bar A_2 -k_2 A_2\bar A_1)}{k_1^2 k_2^2 (k_1 +k_2)^4} ,
\label{ssIJ}\\
&
K=  6(k_1-k_2) (iA_2\bar A_1 - iA_1\bar A_2)
\end{align}
Note this solution \eqref{ss2solitonG}--\eqref{ssIJ} contains a pair of 
arbitrary phases and positions corresponding to the parameters 
\EQ
A_1=|A_1|\exp(i\psi_1) ,\quad
A_2=|A_2|\exp(i\psi_2) 
\endEQ

We now simplify the form of the $2$-soliton solution 
\eqref{ss2solitonG}--\eqref{ss2solitonF} and \eqref{sstheta}--\eqref{ssIJ} 
by the following steps. 
First we observe 
\begin{align}
&
k_2^2 C_1/A_2 = k_1^2 C_2/A_1 = |A_1| |A_2|C
\\
&
J=(C_1/A_2)(\bar C_2/\bar A_1) = (|A_1||A_2||C|)^2/(k_1 k_2)^2
\end{align}
where
\EQ
C= \frac{(k_1-k_2)}{(k_1 +k_2)^2} (k_1 \exp(i\Delta\psi) -k_2 \exp(-i\Delta\psi)) ,\quad
\Delta\psi=\psi_1 -\psi_2
\label{ssC}
\endEQ
Next we introduce the moving coordinates
\EQ
\xi_1 = x-k_1^2 t -x_1 ,\quad
\xi_2 = x-k_2^2 t -x_2 
\endEQ
centered at initial positions
\EQ
x_1 = \ln\big(k_1/(|A_1|\sqrt{|C|})\big)/k_1 ,\quad
x_2 = \ln\big(k_2/(|A_2|\sqrt{|C|})\big)/k_2 
\endEQ
which are determined in terms of $|A_1|$ and $|A_2|$. 
Then the exponential polynomials 
\eqref{ss2solitonG} for $G$ and \eqref{ss2solitonG} for $F$
are given by 
\begin{align}
G=& 
\frac{1}{\sqrt{|C|}} \Big(
k_1\exp(\theta_1)\big( \exp(i\psi_1)+\exp(i(\psi_2+\psi))\exp(2\theta_2) \big)
\nonumber\\&\qquad
k_2\exp(\theta_2)\big( \exp(i\psi_2)+\exp(i(\psi_1+\psi))\exp(2\theta_1) \big)
\Big)
\label{sssimpG}\\
F=&
1+ \exp(2\theta_1 +2\theta_2) +\frac{1}{|C|} \Big(
\exp(2\theta_1) +\exp(2\theta_2) 
\nonumber\\&\qquad
+4 k_1 k_2 (k_1+k_2)^{-2}\big(\exp(i\Delta\psi)+\exp(-i\Delta\psi)\big)
\exp(\theta_1 +\theta_2) 
\Big)
\label{sssimpF}
\end{align}
in terms of 
\EQ
\theta_1=k_1 \xi_1 , \quad
\theta_2=k_2 \xi_2 
\endEQ
and
\EQ
\exp(i2\psi) = C/\bar C 
= \frac{k_1 \exp(i\Delta\psi) -k_2 \exp(-i\Delta\psi)}{k_1 \exp(-i\Delta\psi) -k_2 \exp(i\Delta\psi)}
\label{ssCphase}
\endEQ
These expressions \eqref{sssimpG} and \eqref{sssimpF} match the form of
the colliding solitary wave solution \eqref{ss2soliton}--\eqref{ssF}
if we write 
\EQ
\psi = \phi_2 -\phi_1 ,\quad
\psi_1 = \phi_1 ,\quad
\psi_2 = \phi_1 -\Delta\psi
\endEQ
and use relations \eqref{ssCphase} and \eqref{ssC} to express
\EQ
\exp(i\Delta\psi) = \frac{1}{\gamma}\big(k_1 \exp(i\psi) +k_2 \exp(-i\psi)\big)
= \exp(-i(\nu_1-\psi))= \exp(-i(\nu_2+\psi))
\endEQ
and
\EQ
|C| = \frac{k_1-k_2}{(k_1+k_2)^2} \frac{k_1^2-k_2^2}{\gamma} 
= \frac{1}{\gamma\rho^2}
\endEQ
where
\EQ
\gamma= \sqrt{k_1^2 + k_2^2+2 k_1 k_2\cos(2\psi)} ,\quad
\rho = \frac{\sqrt{k_1+k_2}}{k_1-k_2}
\endEQ

\end{document}